\documentclass[smallextended]{svjour3}     % onecolumn (second format)

\smartqed  % flush right qed marks, e.g. at end of proof
\usepackage{graphicx}
\usepackage{amssymb}
\usepackage{amsmath}
\begin{document}

\title{Gravitational Lensing by Black Holes%\thanks{Grants or other notes
}


\author{Valerio Bozza
}


\institute{V. Bozza \at Dipartimento di Fisica ``E. R.
Caianiello", Universit\`a di
Salerno, I-84084 Fisciano, Italy. \\
 Istituto Nazionale di Fisica Nucleare, Sezione di Napoli,
 Italy.\\
 Istituto Internazionale per gli Alti Studi Scientifici, I-84019,
 Vietri sul Mare, Italy.
              \email{valboz@sa.infn.it}           %  \\
 }

\date{Received: date / Accepted: date}

\maketitle
%
%\begin{abstract}
%Insert your abstract here. Include keywords, PACS and mathematical
%subject classification numbers as needed.
%\end{abstract}

\section{Introduction}

Whenever bending of light by gravitational fields is detected in
astronomical observations it is generally very weak. Even in the
spectacular formation of giant arcs, typical of the so-called
strong lensing, photons are deflected by no more than a few
arcseconds. This leads to the conclusion that the first order
Einstein formula for the deflection angle is sufficient to explain
all observed phenomenology% \cite{KliFri}
. On the other hand, the existence of very compact objects, such
as neutron stars or black holes, is now well-established through
many independent astrophysical observations. In the neighbourhood
of such objects, electromagnetic radiation is generated and
travels through very strong gravitational fields. In such extreme
cases, the calculation of the deflection of photons needs to be
pushed beyond the first order approximation on which the Einstein
formula relies. It must be said that no direct observation of
gravitational lensing by black holes or other compact objects has
ever been performed up to date. Yet, some peculiarities of the
emission spectra of black holes, such as the shape of the
$K_\alpha$ line of the iron, have been interpreted as the final
effect of the strong deflection of photons emitted by the
accretion disk.

The aim of this article is to review the theoretical aspects of
gravitational lensing by black holes, and discuss the perspectives
for realistic observations. We will first treat lensing by
spherically symmetric black holes, in which the formation of
infinite sequences of higher order images emerges in the clearest
way. We will then consider the effects of the spin of the black
hole, with the formation of giant higher order caustics and
multiple images. Finally, we will consider the perspectives for
observations of black hole lensing, from the detection of
secondary images of stellar sources to the interpretation of iron
K-$\alpha$ lines and direct imaging of the shadow of the black
hole.

\section{Spherically Symmetric Black Holes}

The first presentation of the exact calculation of light
deflection by a compact body using the Schwarzschild metric dates
back to the work by Charles Darwin\footnote{Charles Darwin was the
grandson of the famous evolutionist with the same name, author of
``The Origin of the Species".} in 1959 \cite{Dar}.

A generic static spherically symmetric metric can be put in the
form
\begin{equation}
ds^2=A(r)dt^2-B(r)dr^2-C(r)(d\vartheta^2+\sin^2\vartheta d\phi^2).
\label{Sph Metric}
\end{equation}
In particular, the Schwarzschild metric has $A(r)=1-2M/r$,
$B(r)=\left[A(r)\right]^{-1}$, $C(r)=r^2$. Suppose that a photon
comes from infinite distance, grazes the black hole at a minimum
distance $r_m$ and goes away to infinity. The deflection angle,
defined as the angle between the asymptotic incoming and outgoing
trajectories, can be easily derived by the analysis of the
geodesics equations \cite{Wei}
\begin{equation}
\alpha = -\pi+2\int\limits_{r_m}^{\infty}
u\sqrt{\frac{B(r)}{C(r)\left[C(r)/A(r)-u^2\right]}}dr,
\label{alpha Sch}
\end{equation}
where $u$ is the impact parameter, defined as the distance between
the black hole and each of the asymptotic photon trajectories,
related to the closest approach distance $r_m$ by
\begin{equation}
u^2=\frac{C(r_m)}{A(r_m)}. \label{u rm}
\end{equation}

The integral (\ref{alpha Sch}) was solved by Einstein assuming
that $r\geq r_m \gg 2M$. In this limit it reduces to the standard
Einstein result for the Weak Deflection Limit (WDL)
\begin{equation}
\alpha_{WDL} = 4M/u, \label{alpha WDL Sch}
\end{equation}
with $u\simeq r_m$. Darwin, instead, made the integration exactly
in the case of a Schwarzschild metric. His deflection angle is
\begin{equation}
\alpha=-\pi+4\sqrt{r_m/s} F(\varphi,m), \label{alpha Darwin}
\end{equation}
where $F(\phi,m)$ is the elliptic integral of the first
kind\footnote{In our notations $F(\varphi,m)=\int\limits_0^\varphi
\frac{d\theta}{\sqrt{1-m \sin^2\theta}}$.}, and
\begin{eqnarray}
&&s=\sqrt{(r_m-2M)(r_m+6M)} \\
&& m=(s-r_m+6M)/2s \\
&& \varphi=\arcsin \sqrt{2s/(3r_m-6M+s)}.
\end{eqnarray}

\begin{figure}
\includegraphics{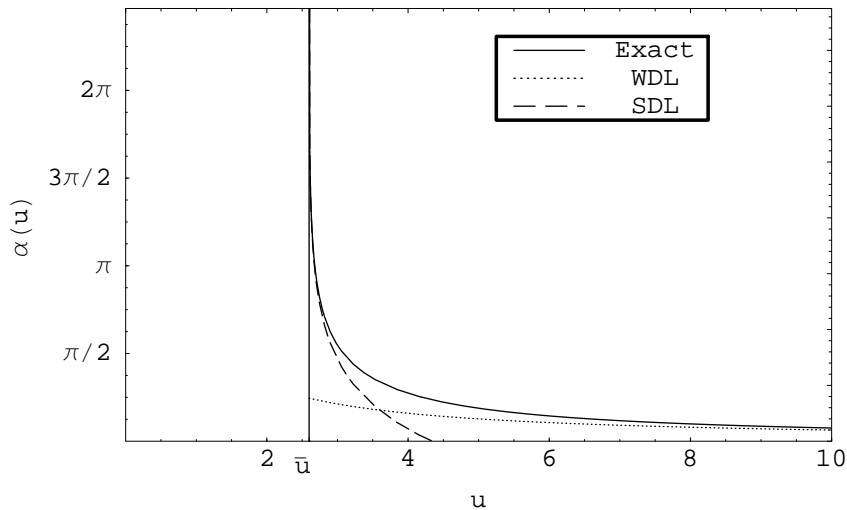}
\caption{Comparison between the exact deflection angle for photons
in a Schwarzschild metric and the two limits discussed in the
text. The impact parameter is in units of the Schwarzschild
radius.} \label{Fig alpha}
\end{figure}

Eq. (\ref{alpha Darwin}) gives the deflection angle as a function
of the unobservable minimum distance $r_m$. We can better express
it in terms of the impact parameter $u$ (which is
coordinate-independent), using Eq. (\ref{u rm}). The exact
deflection angle is plotted in Figure \ref{Fig alpha}, along with
its approximations in the WDL (\ref{alpha WDL Sch}), and the
Strong Deflection Limit (SDL) to be introduced below (\ref{alpha
SDL Sch}).

Darwin's deflection angle and the Einstein approximation coincide
at very large impact parameters. At small impact parameters,
however, the salient feature of the exact deflection angle is the
divergence at a finite value $u=\bar u=3\sqrt{3}M$, corresponding
to a closest approach $r_m=\bar r=3M$ via Eq. (\ref{u rm}). Light
rays with impact parameters closer and closer to $\bar u$
experience larger and larger deflections even exceeding $2\pi$.
This simply means that a light ray deflected with $\alpha$ in the
interval $[2n\pi,2(n+1)\pi]$ performs $n$ complete loops around
the black hole before leaving it. A light ray with impact
parameter exactly equal to $\bar u$ would perform an infinite
number of loops approaching $r=\bar r$ indefinitely. Nonetheless,
circular orbits of light rays with radius $r=\bar r$ are admitted
by geodesics equations, but are unstable against small
perturbations, which would finally drive photons into the black
hole or toward spatial infinity. $\bar r$ is also called the
radius of the photon sphere (for a general definition see Ref.
\cite{CVE}). Finally, photons with impact parameters smaller than
$\bar u$ are just captured by the black hole and fall into the
horizon.

A good approximation for the deflection angle near the divergence
is obtained by expanding the elliptic integral for $r_m \simeq
\bar r$. We get
\begin{equation}
\alpha_{SDL}=-\log (u/\bar
u-1)+\log\left[216(7-4\sqrt{3})\right]-\pi, \label{alpha SDL Sch}
\end{equation}
which is the Strong Deflection Limit approximation plotted in Fig.
\ref{Fig alpha}.

Darwin himself intuited the possibility that a source lensed by a
compact ``star'' with radius smaller than $\bar r$ would be
replicated in two infinite sequences of images on each side of the
lens\footnote{The existence of higher order images and/or the SDL
expansion of Eq. (\ref{alpha SDL Sch}) have been independently
re-discovered several times \cite{Atk,Oha,VirEll,BCIS},
demonstrating how easy is to lose memory of old papers.}. In fact,
thanks to the divergence of $\alpha$, we can always find light
rays with an impact parameter $u$ close enough to $\bar u$ so as
to perform $n$ loops around the black hole (in either way) and
finally reach the observer. These $n$-loop images or higher order
images (called ghosts by Darwin and sometimes mirages or
relativistic images) are missed in the WDL but can be
well-described using the SDL approximation for the deflection
angle.

In the following subsections we will discuss the black hole lens
equation and derive the general features of the images. Finally we
will review the main approximation schemes (WDL, SDL and
non-perturbative).

\subsection{The Black Hole Lens equation}

A lens equation is a relation between the geometry (positions of
source, observer and lens in a given coordinate system) and the
position of the images in the observer's sky. Solving this
equation, we can get the positions of the images as functions of
the geometric parameters of the system.

The position of the observer and the final direction of the photon
hitting the observer can be considered as initial conditions for
the geodesics equations. Tracing them backward in the affine
parameter, we can write down a formally exact lens equation
\cite{FriNew}. This approach has been applied to the Schwarzschild
lens \cite{FKN} and then generalized to all spherically symmetric
spacetimes \cite{Perlick} (see also \cite{LivRev}).

Let us consider a source at coordinates\footnote{The spherical
symmetry ensures that the motion takes place on a fixed plane.
Choosing our coordinate system so that
$\vartheta_O=\vartheta_S=\pi/2$ we get $\vartheta=\pi/2$ during
the whole trajectory.} $(r_S,\phi_S)$ and an observer at
$(r_O,\phi_O)$, with $r_O>r_S$. In typical lensing situations, the
photon is emitted from the source, approaches the black hole to a
minimum distance $r_m$ and then escapes toward the observer.
Therefore, we can divide the trajectory into an approach phase
(from $r_S$ to $r_m$ with\footnote{The dot denotes derivative with
respect to the affine parameter.} $\dot r<0$) and an escape phase
(from $r_m$ to $r_O$ with $\dot r>0$). By simple integration of
the geodesics equations, we can calculate the total azimuthal
shift experienced by the photon as
\begin{equation}
\Phi_-(u,r_{O},r_{S})=\left[\int\limits_{
r_m}^{r_{O}}+\int\limits_{r_m}^{r_{S}}\right] u
\sqrt{\frac{B(r)}{C(r)\left[C(r)/A(r)-u^2\right]}}dr. \label{Phi-}
\end{equation}
The two integrals separately cover the approach and escape phases
of the trajectory. $u$ is the asymptotic impact parameter, related
to $r_m$ by Eq. (\ref{u rm}).

However, we must also consider the possibility that the photon
directly goes from the source to the observer always moving away
from the black hole with $\dot r>0$ (e.g. when the source is
located between the lens and the observer). In this case, there is
only an escape phase and the azimuthal shift is
\begin{equation}
\Phi_+(u,r_{O},r_{S})=\int\limits_{ r_S}^{r_{O}} u
\sqrt{\frac{B(r)}{C(r)\left[C(r)/A(r)-u^2\right]}}dr. \label{Phi+}
\end{equation}
Note that the two expressions coincide when $r_{S}\rightarrow
r_m$, i.e. when the photon is emitted in a tangential direction
$\dot r=0$.

The lens equation can be written by simply imposing that the
azimuthal shift equals the difference in azimuth between source
and observer modulo $2\pi$.
\begin{equation}
\Phi(u,r_{O},r_{S})=\Delta\phi\equiv \phi_O-\phi_S+2n\pi,
\label{ExactLE}
\end{equation}
where $\Phi=\Phi_-$ if $\Delta\phi>\Phi_\pm(u,r_{O},r_m)$ and
$\Phi=\Phi_+$ otherwise. The integer number $n$ takes into account
the fact that $\phi$ is a periodic coordinate.

The angle $\theta$ at which the observer detects the photon is
\cite{Nem}
\begin{equation}
\theta=\arcsin \left(u\sqrt{\frac{A(r_{O})}{C(r_{O})}} \right).
\label{thetau}
\end{equation}

Once we fix the radial coordinates of source $r_S$ and observer
$r_O$, and their relative azimuthal position $\Delta\phi$, by
inverting the lens equation (\ref{ExactLE}) we can get $u$ and
thus the positions of the images $\theta$ through Eq.
(\ref{thetau}).

\begin{figure*}
\resizebox{10cm}{!}{\includegraphics{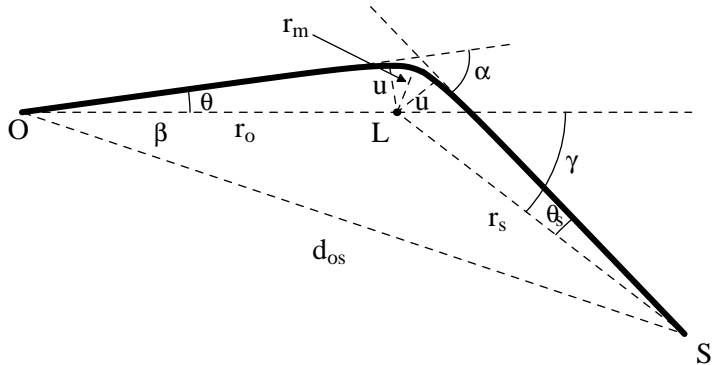}} \caption{Generic
lensing geometry by a static spherically symmetric body $L$,
illustrating the quantities defined in the text. The source is in
$S$ and the observer in $O$.} \label{Fig LEq}
\end{figure*}

An interesting limit is obtained when source and observer are very
far from the lens with respect to all other distance scales ($u$
and $M$). Taking only the smallest powers in $u/r_O$ and $u/r_S$,
we get the Ohanian lens equation \cite{Oha}
\begin{equation}
\alpha-\theta-\theta_S=-\gamma\equiv\Delta \phi-\pi, \label{LensEQ
Oha}
\end{equation}
with $\sin\theta\simeq u/r_O$, $\sin\theta_S=u/r_S$ and $\alpha$
being the deflection angle between the asymptotic incoming and
outgoing trajectories ($\alpha=\Phi(u,\infty,\infty)-\pi$, see
Fig. \ref{Fig LEq}). The Ohanian lens equation is an appealing
simplification of Eq. (\ref{ExactLE}), in that it gets rid of the
two parameters $r_O$ and $r_S$. While the limit $r_O \simeq
\infty$ is justified from the fact that the Earth is very far from
any known candidate black hole, the limit $r_S \gg u$ leaves out
the very interesting case in which the source is part of the
environment of the black hole. A discussion on the errors
committed when using the Ohanian or other approximate lens
equations appeared in the literature is contained in Ref.
\cite{ComLEq}.

\subsection{Position of the images}

Armed with our exact lens equation (\ref{ExactLE}), we can now
find the images for a generic source position varying from $\Delta
\phi=0$ (source in front of the black hole) to $\Delta\phi=\pi$
(source behind the black hole). Only images on the same side of
the source (direct images) can be obtained by Eq. (\ref{ExactLE}),
since we have assumed that $\dot\phi$ is always positive. However,
thanks to the spherical symmetry, the position of the images on
the opposite side of the source (indirect images) can be obtained
by solving Eq. (\ref{ExactLE}) with $\Delta\phi \rightarrow
2\pi-\Delta \phi$. Thanks to the $2n\pi$ on the right hand side of
Eq. (\ref{ExactLE}), we have one image in each interval
$(m-1)\pi<\Phi<m\pi$ for any integer $m$. We will refer to this
integer as the order of the image. The primary image will thus
have $0<\Phi<\pi$, the secondary image $\pi<\Phi<2\pi$ and so on
for higher order images. Odd order images are direct and even
order images are indirect.

\begin{figure*}
\resizebox{\hsize}{!}{\includegraphics{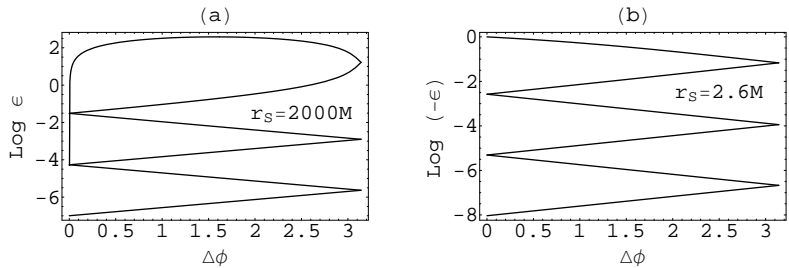}}
\caption{Position of the images as a function of the source
position $\Delta \phi$ for $r_S=2000M$ (a) and $r_S=2.6M$ (b). In
both cases $r_O=\infty$. $\epsilon\equiv(\theta/\bar \theta -1)$
is the relative distance of the image from the shadow border.}
\label{Fig Images}
\end{figure*}

In Fig. \ref{Fig Images}, we plot the position of the images as a
function of $\Delta \phi$. The left panel, represents the case of
a distant source ($r_S=2000M$), while the right panel considers a
source very close to the black hole between the horizon $r=2M$ and
the photon sphere $r=\bar r=3M$. The position of the images is
expressed in terms of $\epsilon\equiv(\theta/\bar \theta -1)$,
where $\bar \theta\equiv \bar u/r_O$ is called the shadow border
for reasons that will be clear in a while. From top to bottom, the
curves represent the primary image ($n=0$), the secondary image
($\Delta \phi \rightarrow 2\pi-\Delta \phi$, $n=0$), the third
order image ($n=1$), the fourth order image ($\Delta \phi
\rightarrow 2\pi-\Delta \phi$, $n=1$), the fifth and sixth order
images ($n=2$).

The familiar gravitational lensing geometry with the source behind
the black hole is recovered at $\Delta \phi=\pi$. In case of
perfect alignment, we expect the primary and secondary images to
merge in the classical Einstein ring. Indeed, in Fig. 1a we see
that they tend to the same value of $\epsilon$, that is to the
same angle $\theta$ (but on opposite sides). As soon as the source
moves out of perfect alignment ($\Delta\phi<\pi$), the primary
image moves farther from the black hole while the secondary image
gets closer. For a source in quadrature ($\Delta\phi\simeq \pi/2$)
the light bending is practically negligible for the direct image,
so $\theta$ reaches the maximum value $r_S/r_O$ and then decreases
to zero when $\Delta\phi=0$ (source in front of the black hole).
On the other hand, the secondary image gets closer and closer to
the black hole but always stays outside the circle of radius
$\bar\theta$.

Coming to higher order images, we see that they also form Einstein
rings when $\Delta\phi=\pi$, but with radius just very slightly
larger than $\bar\theta$. As $\Delta\phi$ decreases, the direct
images move farther and the indirect get closer to the shadow
border. Without the logarithmic scale, all images would appear
packed together at the shadow border $\bar\theta$. Finally, when
the source is in front of the black hole ($\Delta\phi=0$), every
indirect image merges with the direct image of the next order to
form an Einstein ring. Therefore, black holes surprisingly behave
in a similar way with sources in front of the black hole and
behind the black hole, since we can have an infinite sequence of
concentric Einstein rings in both cases. The geometry with the
source in front of the black hole is also known as retro-lensing
\cite{Lum,HolWhe}. It is important to stress that in no case
images of sources outside the photon sphere can appear inside the
circle of radius $\bar\theta$, which justifies its name as shadow
of the black hole.

Now let's move to Fig. \ref{Fig Images}b, which represents the
same diagram for a source inside the photon sphere but outside the
horizon. In this case, $\epsilon=\theta/\bar\theta-1$ is always
negative, which means that the images form inside the shadow. The
distance order of the images from the black hole is now reversed,
with lower order images being closer to the black hole and higher
order images appearing closer and closer to the border of the
shadow. Apart from this major difference, the diagram of Fig.
\ref{Fig Images}b is quite similar to \ref{Fig Images}a, with the
formation of Einstein rings at $\Delta\phi=0,\pi$. Just note that
at $\Delta \phi=0$ the direct image is again in $\theta=0$,
corresponding to $\epsilon=-1$.

The dependence of the images on the source distance is illustrated
in detail in Fig. \ref{Fig Images rS}. This figure shows how the
radius of the Einstein rings (in standard lensing configuration
$\Delta\phi=\pi$) depends on the source position. In Fig. \ref{Fig
Images rS}a we have the primary Einstein ring and in Fig. \ref{Fig
Images rS}b we have the second order Einstein ring. In both cases
the rings enter the shadow when $r_S<3M$. At very large $r_S$,
however, the primary Einstein ring grows according to the WDL
formula $\theta_E=\sqrt{4M r_S/r_O(r_O+r_S)}$, whereas the higher
order Einstein rings stay close to the shadow border.

\begin{figure*}
\resizebox{\hsize}{!}{\includegraphics{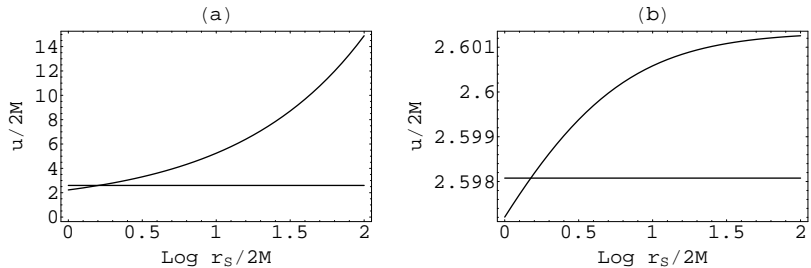}}
\caption{Position of the Einstein ring vs source distance. (a)
Primary ring, (b) Second order ring. The horizontal line
represents the shadow border $\bar u=3\sqrt{3}M$.} \label{Fig
Images rS}
\end{figure*}

\subsection{Magnification}

The magnification of the images only makes sense for sources in
the asymptotically flat region $r_S \gg M$. In fact, in order to
compare angular elements of image and source, we need to project
the source element to a Minkowski space. This process is
coordinate-dependent unless the source is in the asymptotic
region. Furthermore, the intensity of the image elements is also
red-shifted if the source is deep in the gravitational field of
the lens as a consequence of the conservation of the number of
photons $I_O/\nu_O^3=I_S /\nu_S^3$.

Let us consider a source far from the black hole. An angular area
element of the image has size $du/r_O$ in the radial direction and
$u d\eta/r_O$ in the tangential direction, where $\eta$ is an
angular coordinate around the axis lens-observer. The
corresponding tangential displacement in the source surface is
$r_S \sin\Delta\phi \; d\eta$. The tangential magnification is
therefore
\begin{equation}
\mu_t=\frac{u/r_O}{r_S\sin\Delta\phi/d_{OS}}, \label{mut}
\end{equation}
where
\begin{equation}
d_{OS}=\sqrt{r_O^2 + r_S^2 -2r_Or_S \cos \Delta\phi}
\end{equation}
is the (Euclidean) distance between source and
observer\footnote{The Euclidean distance can be safely defined for
sources and observers in the asymptotic region $r_S,r_O \gg M$,
otherwise it loses meaning.}.

For the radial displacement the situation is more subtle. A shift
in the impact parameter $du$ selects a geodesic close to the
original one. By differentiation of the lens equation
(\ref{ExactLE}), we get
\begin{equation}
\frac{\partial \Phi}{\partial u}du + \frac{\partial \Phi}{\partial
r_S}dr_S=-d\phi_S. \label{dPhi}
\end{equation}

This equation approximates the perturbed geodesic in a
neighbourhood of the source ($r_S$,$\phi_S$). The point in which
this geodesic will intercept a section orthogonal to the emission
direction $\dot x_e^i$ is $(r_S,\phi_S)+dy^i$ with
\begin{equation}
dy^i\equiv (dr_S,d\phi_S)=
\left(u/r_S,\mp\sqrt{1-u^2/r_S^2}/r_S\right) dR, \label{dy}
\end{equation}
which satisfies\footnote{Here
$\eta_{ij}=diag(1,r^2,r^2\sin^2\vartheta)$ is the Euclidean metric
in polar coordinates as the source is in the asymptotic region.}
$\eta_{ij}\dot x_e^idy^j=0$ and $\eta_{ij}dy^idy^j=dR^2$. The
distance $dR$ spans a source element. We can thus define the
angular source element in absence of gravitational lensing as
$dR/d_{OS}$. The double sign in $d\phi_S$ in Eq. (\ref{dy})
follows the sign of $\Phi_\pm$ that applies to the image under
examination.

Finally, the radial magnification is
\begin{equation}
\mu_r=\frac{du/r_O}{dR/d_{OS}},
\end{equation}
where $du$ and $dR$ are related by Eqs. (\ref{dPhi}) and
(\ref{dy}). The total magnification is $\mu=\mu_r\mu_t$.

When $u \ll r_S$ and the photon is emitted with $\dot r<0$, our
definition of magnification reduces to \cite{Oha}
\begin{equation}
\mu|_{r_S\rightarrow\infty}=\frac{d_{OS}^2}{r_S^2}
\frac{\sin\theta}{\sin \Delta\phi} \frac{d\theta}{d \Delta\phi}.
\end{equation}

\begin{figure*}
\resizebox{8cm}{!}{\includegraphics{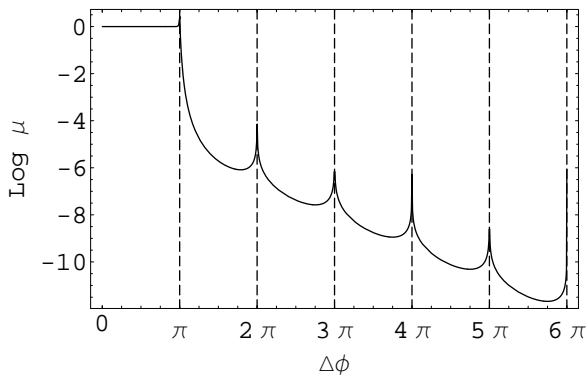}}
\caption{Magnification of the images as a function of the
azimuthal shift $\Delta \phi$ for the Schwarzschild lens.}
\label{Fig mag}
\end{figure*}

Fig. \ref{Fig mag} shows the magnification as a function of the
the azimuthal shift \cite{Oha}. All images have been put in
sequence by adding $(m-1)\pi$ to the abscissa of images of order
$m$. We see that the magnification diverges at all multiples of
$\pi$. This divergence comes from the tangential magnification
(\ref{mut}), which means that images of a finite size-source
become large arcs and then merge into rings when the source is in
front or behind the black hole. The primary image
($0<\Delta\phi<\pi$) has otherwise magnification very close to 1,
whereas all other images have an exponentially decreasing
magnification. The secondary image interpolates between the two
behaviours. We conclude that the lines defined by $\Delta\phi=0$
and $\Delta\phi=\pi$ and extending from the horizon to infinity
($2M<r<\infty$) represent degenerate (with zero section) caustic
tubes for the Schwarzschild metric.

The parity of the images is determined by the factor
$\sin\Delta\phi$ in the tangential magnification. All direct
images have positive parity and all indirect images have negative
parity.

We conclude by saying that possible effects beyond the geometrical
optics scheme for photons travelling close to the photon sphere
are still awaiting a full investigation. The existence of
nonlinear dispersion relations for the photons propagating close
to the photon sphere could have important consequences not yet
considered \cite{DecFol}.

\subsection{Analytical approximations}

The elliptic integrals of the exact treatment of light bending are
not particularly illuminating and prevent an analytical inversion
of the lens equation (\ref{ExactLE}). For this reason, several
approximation schemes have been developed in different limits.
With their analytical formulae they shed new light on the plots of
the previous subsection. Furthermore, taking advantage of the
perturbative schemes in different limits, it is possible to
compare different black hole metrics coming from alternative
gravity theories and use gravitational lensing to discriminate
among different scenarios.

\subsubsection{Weak deflection limit}

The WDL amounts to setting $u\gg M$ in the deflection integrals
(\ref{alpha Darwin}) or (\ref{Phi-}) and (\ref{Phi+}), which also
implies $r_m,r_S\gg M$. We are thus in the lower-right corner of
Fig. \ref{Fig alpha} or the upper side of Fig. \ref{Fig Images}a.
When $\Delta\phi$ is significantly less than $\pi$, the
gravitational lensing effect on the primary image is negligible
while the secondary image is not in the WDL regime. Therefore, the
WDL approximation is only useful in the case $\Delta\phi\simeq
\pi$, where it can be nicely employed to describe the primary and
secondary images.

In the WDL it is possible to study a generic static spherically
symmetric metric by using the PPN formalism. In isotropic
coordinates, the metric can be parameterized as
\begin{eqnarray}
&& ds^2=A'(r')dt^2-B'(r')\left[{dr'}^2+{r'}^2d\Omega^2 \right] \\
&& A'(r')=1-2\alpha' \frac{M}{r'}+2\beta' \frac{M^2}{{r'}^2} +\ldots\\
&& B'(r')=1+2\gamma' \frac{M}{r'}+\frac{3}{2}\delta'
\frac{M^2}{{r'}^2} +\ldots
\end{eqnarray}
In the Schwarzschild limit, the PPN coefficients become
$\alpha'=\beta'=\gamma'=\delta'=1$.

The integrals in the azimuthal shift (\ref{Phi-}) can be easily
expanded in powers of $M$. In particular, the deflection angle
(\ref{alpha Sch}) has been calculated to the second order in $M$
by Epstein and Shapiro already in 1980 \cite{EpsSha} (see also
\cite{WDLSph}). Defining the Einstein angle
\begin{equation}
\theta_E=\sqrt{4M r_S/(r_O d_{OS})},
\end{equation}
a convenient expansion parameter is \cite{WDL}
\begin{equation}
\varepsilon=\theta_E d_{OS}/(4r_S). \label{varepsilon}
\end{equation}

Without gravitational lensing, the observer would see the source
at angle $\beta$ from the lens. Simple geometry relates this angle
to the azimuthal shift $\Delta\phi$ as $d_{OS}\sin \beta=r_S \sin
\Delta\phi$. Setting $y\equiv \beta/ \theta_E$, it turns out that
to lowest order $\theta\sim \theta_E$ and we can expand the
position of the images as $\theta=\theta_E(\theta_0+\varepsilon
\theta_1+\varepsilon^2 \theta_2 +\ldots)$.

Expanding the exact lens equation (\ref{ExactLE}), we can collect
all terms according to their order in $\varepsilon$. Finally, we
can calculate the position of the images and their magnification
$\mu=(\mu_0+\varepsilon \mu_1 +\ldots)$ order by order. We just
quote here the final results\footnote{In these formulae $y$ is
positive for the direct image and negative for the indirect
image.} found by Ebina et al. \cite{Ebina} in the Schwarzchild
case, by Sereno \cite{SerRei} for Reissner-Nordstr\"om and by
Keeton and Petters for a generic PPN metric \cite{WDL}
\begin{eqnarray}
&& \theta_0=\left(y+\sqrt{y^2+A_1}\right)/2, \;\;
\theta_1=A_2/(A_1+4\theta_0^2) \\
&& \mu_0=16\theta_0^4/(16\theta_0^4-A_1^2), \; \; \mu_1=-
16\theta_0^3A_2/(A_1+4\theta_0^2)^3
\end{eqnarray}
with the coefficients $A_1$ and $A_2$ fixed by the PPN metric as
\begin{equation}
A_1=2(\alpha'+\gamma'), \; \;
A_2=2{\alpha'}^2-\beta'+2\alpha'\gamma'+3\delta'/4.
\end{equation}
In principle, if the source is close enough to the lens so as to
raise $\varepsilon$ significantly, by studying gravitational
lensing we can gain access to the first order correction in the
WDL expansion and thus infer the values of PPN parameters.
Surprisingly, it turns out that all observables that are
independent of the unknown source position are unchanged at first
order in $\varepsilon$ \cite{Ebina} for all PPN metrics agreeing
with General Relativity at zero order (thus having
$\alpha'+\gamma'=2$) \cite{WDL}. This fact was prefigured by
Damour and Esposito-Far\`ese \cite{DamEsp} using an independent
field theory approach to gravity theories in which they showed
that 2PN parameters do not enter the equation of motion of light.

Finally, we note that the exact lens equation (\ref{ExactLE}) is
not really necessary to calculate the first order correction, as
one would reach the same result using the classical small angles
lens equation with a second order expansion of the deflection
angle (\ref{alpha Sch}). The second order corrections in
$\varepsilon$, however, can only be calculated by the exact lens
equation or by the Ohanian lens equation (\ref{LensEQ Oha})
\cite{SerCos}, as other lens equations would deviate from the
correct results. The Ohanian lens equation, instead, would fail at
the fourth order only \cite{ComLEq}.

\subsubsection{Strong Deflection Limit}

The expansion of the elliptic integral in Darwin's formula at its
divergence at $r=3M$ leads to the SDL formula (\ref{alpha SDL
Sch}), which shows that the divergence at $u=3\sqrt{3}M$ is
logarithmic \cite{Dar,BCIS,Lum,Cha}. This is not a peculiarity of
Schwarzschild metric, as even in Reissner-Nordstr\"{o}m metric a
similar divergence can be found \cite{ReiNor}. Actually, it is
possible to prove that the logarithmic divergence in the
deflection angle is a universal feature of all static spherically
symmetric metrics endowed with a photon sphere
\cite{Boz1}\footnote{It is easy to prove that all static
spherically symmetric asymptotically flat metrics with an event
horizon have a photon sphere.}.

In order to calculate the SDL in a given metric, it is first
necessary to calculate the radius of the photon sphere. This is
obtained by imposing that the denominator of Eq. (\ref{alpha Sch})
has a double root at $r=r_m$. This is equivalent to the equation
\begin{equation}
C'(\bar r)/C(\bar r)=A'(\bar r)/A(\bar r),
\end{equation}
which yields the radius of the photon sphere $\bar r$ \cite{CVE}.
The minimum impact parameter $\bar u$ is then given by Eq. (\ref{u
rm}) with $r_m=\bar r$.

We can then expand the integrals in the azimuthal shift
(\ref{Phi-}) in terms of $\epsilon=(u/\bar u -1)$ to get the SDL
lens equation\footnote{The SDL was developed in Ref. \cite{Boz1}
for standard lensing only. The extension to the retro-lensing
configuration was done in Ref. \cite{EirTor}. Here we present the
most general treatment of Ref. \cite{SDLDLS}.}
\begin{equation}
\Delta\phi+2n\pi=-\bar a \log [\epsilon/(1-\bar r/r_S)]+\bar b,
\label{LE SDL}
\end{equation}
where the coefficients $\bar a$ and $\bar b$ depend on the metric
considered\footnote{They can be calculated analytically or
numerically using the procedure described in Ref. \cite{Boz1} for
sources at infinity and in Ref. \cite{SDLDLS} for any source
positions.} and on the source radial coordinate $r_S$. Indeed, the
SDL coefficients may vary in different alternative theories of
gravity, representing a sort of identity card of the black hole
metric and a possible key for discriminating between theories
predicting deviations from General Relativity in strong fields
\cite{Spheric,BozRev}. In the Schwarzschild case
\begin{equation}
\bar a=1, \;\;\;\; \bar b=
-2\log\left[\left(3+\sqrt{3}\right)\left(3+\sqrt{3+18M/r_S}\right)/(36\sqrt{6})\right].
\label{FinDLS}
\end{equation}
Corrections to Eq. (\ref{LE SDL}) at higher orders in $\epsilon$
have been worked out only in the Schwarzschild case for
$r_S\rightarrow \infty$ \cite{IyePet}.

This lens equation can be used to describe all images with $n\geq
1$, generated by photons looping at least once around the black
hole (cfr. Fig. \ref{Fig Images}a). It becomes inaccurate only for
the secondary image. It may even be used for sources inside the
photon sphere (Fig. \ref{Fig Images}b). The position of the images
is found by inverting Eq. (\ref{FinDLS}) and remembering that
$\theta=u/r_O$ in the limit $r_O \gg M$. We have \cite{SDLDLS}
\begin{equation}
\theta_n=\bar \theta \left[1+(1-\bar r/r_S) e^{(\bar
b-\Delta\phi-2n\pi)/\bar a} \right].
\end{equation}
The indirect images are found by replacing $\Delta\phi$ by
$2\pi-\Delta\phi$ as usual. This formula indicates that the higher
order images tend to the shadow border $\bar\theta$ with an
exponential law (which justifies our choice of the logarithmic
scale in Fig. \ref{Fig Images}). The images form outside the
border for $r_S > \bar r$ and inside it otherwise.

For sources very far from the black hole, the magnification is
\cite{Oha,Boz1}
\begin{equation}
\mu_n= \left(\frac{D_{OS}}{r_S}\right)^2\frac{\bar \theta^2}{\bar
a \sin \Delta\phi}e^{(\bar b-\Delta\phi-2n\pi)/\bar a},
\label{mun}
\end{equation}
which nicely reproduces the behavior of Fig. \ref{Fig mag} and
also proves that higher order images are exponentially fainter and
fainter.

It is interesting to note that the time delay between two
consecutive higher order images is approximately $\Delta T=2\pi
\bar \theta r_O$, which could be used in principle to infer the
mass and distance of the black hole \cite{TimDel}.

Another interesting application of the SDL scheme is in the study
of a static spherically symmetric black hole embedded in an
external gravitational field. In the case of the standard WDL,
Chang and Refsdal \cite{ChaRef} showed that the zero order
Einstein ring becomes elliptical and the corresponding degenerate
caustic tube at $\Delta\phi=\pi$ acquires a diamond-shaped
cross-section with angular size proportional to the external shear
$\gamma$ and $\theta_E$. In the SDL, a correct description of the
tidal forces \cite{Poi} leads to the conclusion that also higher
order Einstein rings gets slightly distorted and their
corresponding caustic tubes are diamond-shaped with size
proportional to $\gamma \bar \theta$ \cite{BozSer}.

\subsubsection{Non-perturbative methods}

While the WDL describes the two main images in the best alignment
regime between source and lens ($\Delta\phi \simeq \pi$) and the
SDL describes all higher order images ($n>0$), no perturbative
approximations are possible for the secondary image when the
source is not aligned behind the black hole ($n=0$ and
$0<\Delta\phi<\pi$). Indeed Figs. \ref{Fig Images} and \ref{Fig
mag} show that the secondary image interpolates between the WDL
and the SDL regimes when $\Delta\phi$ runs from $\pi$ to 0, going
from $\theta\sim\theta_E$ to $\theta \sim\bar\theta$.

A nice strategy to obtain simple analytic formulae has been
proposed by Amore and collaborators \cite{Amore}. First, the
integral (\ref{alpha Sch}) is written in the form
\begin{equation}
\alpha=2\int\limits_0^1\frac{dz}{\sqrt{V(1)-V(z)}}-\pi
\label{alphaAmore}
\end{equation}
with $z=r_m/r$ and
\begin{equation}
V(z)=z^2 C(r)/B(r)-C^2(r)A(r_m)/[B(r)A(r)C(r_m)]+A(r_m)/C(r_m)
\end{equation}
being a sort of potential\footnote{Interestingly, the lensing
problem is reformulated as an oscillator problem.} in the variable
$z$.

Then, the potential $V(z)$ is replaced by a simpler form $V_0$ and
the integral becomes
\begin{eqnarray}
&& \alpha= 2\int\limits_0^1\frac{dz}{\sqrt{V_0(1)-V_0(z)}}
\frac{1}{\sqrt{1+\delta\Delta(z)}} - \pi \label{alphadeltaAmore}
\\
&& \Delta(z)=\frac{V(1)-V(z)}{V_0(1)-V_0(z)}-1.
\end{eqnarray}
Finally, the integral is solved as a series in powers of $\delta$
and calculated in $\delta=1$ (where Eq. (\ref{alphadeltaAmore})
coincides with (\ref{alphaAmore})). The convergence of the series
depends on the chosen form of the auxiliary potential $V_0$. If
$V_0$ depends on some shape parameter $\lambda$, this can be fixed
by requiring the minimum sensitivity of the deflection angle (as
in a variational method).

As a practical example, taking $V_0=\lambda z^2$ in the
Schwarzschild metric, Amore and Arceo \cite{Amore} obtained
\begin{equation}
\Delta\phi=\pi\left(\frac{1}{\sqrt{1-8M/(\pi r_m)}}-1\right),
\end{equation}
which works nicely in the intermediate regime for the zero order
indirect image. More sophisticated potentials may lead to even
better approximations englobing the main features of the WDL and
SDL \cite{Amore} at the price of dealing with more involved
formulae.

\section{Rotating black holes}

Most astrophysical systems are known to have non-negligible
angular momentum. Therefore it is natural to expect that physical
black holes should be properly described by the Kerr metric. The
study of null geodesics in Kerr geometry has been initiated by
Carter in 1968 \cite{Car}, who separated the Hamilton-Jacobi
equation and wrote the geodesics equations in terms of first
integrals of the motion
\begin{eqnarray}
&& \pm \int \frac{dr}{\sqrt{R}}=\pm \int \frac{d
\vartheta}{\sqrt{\Theta}} \label{Geod1}\\
& \phi_f-\phi_i =& a \int\frac{r^{2}+a^{2}-a J}{\Delta \sqrt{R}}
dr-a \int \frac{dr}{\sqrt{R}} + J \int
\frac{\csc^2\vartheta}{\sqrt{\Theta}} d \vartheta, \label{Geod2}
\end{eqnarray}
where
\begin{eqnarray}
&\Theta=&Q+a^2 \cos^2\vartheta-J^2 \cot^2\vartheta; \;\; \Delta=r^2-2Mr+a^2 \\
&R=&r^4+(a^2-J^2-Q)r^2+2M(Q+(J-a)^2) r -a^2 Q, \label{R}
\end{eqnarray}
$J$ and $Q$ are constants of motion labelling the geodesic under
examination\footnote{$J$ is the projection of the angular momentum
of the photon on the rotation axis, $Q$ is a combination of the
square of the other components of the angular momentum of the
photon with the black hole angular momentum.}, $a$ is the angular
momentum of the black hole divided by its mass and expressed in
natural units. The Kerr metric describes a rotating black hole if
$a<M$ with horizon at $r_H=M+\sqrt{M^2-a^2}$, otherwise it
represents a naked singularity. Both $\dot r$ and $\dot \vartheta$
may change sign during the motion. Therefore, every integral
should be calculated from the emission point to the first
inversion point (if any), then between each pair of consecutive
inversion points, up to the final observation point. In the
following, we will use the notation $\mu=\cos \vartheta$
\footnote{Not to be confused with the magnification $\mu$ of the
previous section.}.

The main difference with respect to spherically symmetric black
holes is that the motion of the photon is no longer confined on a
single plane, as the angular momentum of the black hole induces an
orbital precession around its rotation axis. As a consequence, the
polar angle $\vartheta$ cannot be eliminated by a simple choice of
coordinates and the problem is fully three-dimensional. In
principle, all integrals can be written in terms of elliptic
integrals (see e.g. \cite{RauBla}). This is not particularly
illuminating for the analytic point of view, but may be useful for
implementing numerical codes for Kerr-geodesics tracing
\cite{KVP,Vie,Fanton,DexAgo}. In particular, the analytic
deflection angle for photons confined in the equatorial plane
takes a reasonably short expression \cite{IyeHan1}.

A thorough presentation of the classification and the properties
of Kerr geodesics is in Ref. \cite{Cha}. In order to investigate
the rotation of the polarization plane of electromagnetic
radiation (not discussed in this text), it is useful to introduce
a tetrad formalism \cite{PinRoe}. As for spherically symmetric
case, several studies of Kerr lensing in the WDL
\cite{EpsSha,WDLKer,AsaKas,SerDeL} and in the SDL
\cite{SDLDLS,BozEq,BozKer} have been carried out. These studies
allow to get interesting analytical approximations in some limits
that we will sometimes recall later.

In the following subsections, we will give a sketch of the main
characteristics of lensing by rotating black holes. We will first
describe the shadow of the black hole, then we will discuss the
caustics and finally the formation of the images.

\subsection{Shadow of the Kerr black hole}

Null geodesics are fully identified by the constants of motion $J$
and $Q$. For a static observer in the asymptotic flat space ($r_O
\gg M$), the position in the sky in which a photon is detected is
completely determined by the values of these constants. In fact,
defining $\mu_O=\cos \vartheta_O$, we have
\begin{eqnarray}
&& \theta_{1}=- J\left(r_{O} \sqrt{1-\mu_O^2}\right)^{-1}, \label{Th1J} \\
&& \theta_{2}=\pm r_{O}^{-1}\sqrt{Q + \mu_O^2\left[ a^2-
J^2/(1-\mu_O^2) \right] },\label{Th2Q}
\end{eqnarray}
where $\theta_1$ is the angular distance from the black hole
center projected orthogonally to the spin direction and $\theta_2$
is the same angular distance projected along the projection on the
spin on the sky. The problem of determining the position of the
images of a given source can thus be reformulated in terms of $J$
and $Q$.

Similarly to the spherically symmetric case, the integrals in the
radial coordinate diverge logarithmically if the argument of the
square root has a double root. This happens when $R$ and $\partial
R/\partial r$ simultaneously vanish at the same point $\bar r$.
The two equations $R=0$ and $R'=0$ can be solved in terms of the
two constants of motion $J$ and $Q$ as functions of $\bar r$, so
as to find those geodesics which asymptotically approach an orbit
at a fixed radius $\bar r$. Explicitly, we have
\begin{eqnarray}
&&\bar J(\bar r)=\frac{\bar r^{2}(\bar r-3M)+a^{2}(M+ \bar
r)}{4M^2a(M-
\bar r)} \label{Jm}\\
&& \bar Q(\bar r)=\frac{\bar r^{3}\left[ 16M^3 a^2-\bar r(\bar
r-3M)^2\right]}{16M^4a^{2}(\bar r-M)^2} \label{Qm}
\end{eqnarray}
which describe a locus in the $(J,Q)$ space parameterized by $\bar
r$. Only positive values of $Q$ describe geodesics approaching
infinity (see Eq. (\ref{Th2Q})). This constrains the possible
values of $\bar r$ into a finite interval $[\bar r_+,\bar r_-]$
defined by the condition\footnote{The third degree equation $Q=0$
has always two roots outside the horizon.} $Q\geq 0$. The two
extreme values $\bar r_\pm$ lead to a vanishing $Q$, which
corresponds to a photon moving on an equatorial orbit. The lower
value $\bar r_+$ corresponds to a photon moving with positive $J$,
i.e. rotating in the same sense with the black hole (prograde
photon). The higher value $\bar r_-$ gives a negative $J$, thus
describing a photon counter-rotating with respect to the black
hole (retrograde photon). In-between, we can have orbits with any
possible inclinations.

From the point of view of the observer at infinity, thanks to Eqs.
(\ref{Th1J}) and (\ref{Th2Q}), all geodesics asymptotically
starting in an orbit at fixed radius $\bar r$, described by Eqs.
(\ref{Jm}) and (\ref{Qm}), also describe a locus in the sky
coordinates $(\theta_1,\theta_2)$. Notice that if the observer is
not on the equator, equatorial orbits will never reach it. Indeed,
the range of values of $\bar r$ that describe orbits ending at the
observer is defined by the condition that the argument of the
square root in Eq. (\ref{Th2Q}) is positive. The new range $[\bar
r_<,\bar r_>]$ is obviously smaller than $[\bar r_+,\bar r_-]$ and
coincides with it if the observer is on the equator.

\begin{figure*}
\resizebox{6cm}{!}{\includegraphics{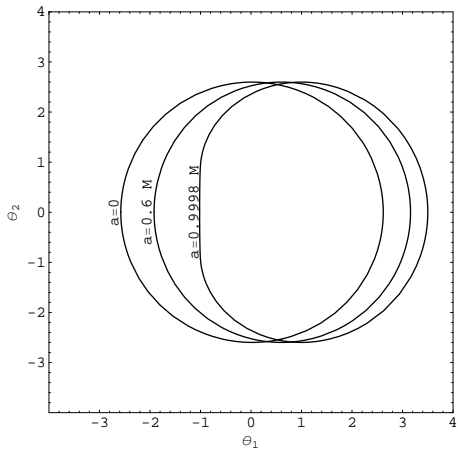}} \caption{The shadow
border as seen by an equatorial observer for different values of
the black hole spin. $\theta_1$ and $\theta_2$ are the angular
coordinates in the observer's sky in units of $2M/r_O$. The black
hole is in the origin and the spin axis points in the positive
$\theta_2$ direction. }
 \label{Fig shadow}
\end{figure*}

Analogously to spherically symmetric black holes, the locus
$(\bar\theta_1(\bar r), \bar \theta_2(\bar r))$ is called the
shadow border of the Kerr black hole. In Fig. \ref{Fig shadow} we
show the shadow border for three values of the spin parameter $a$.
Indeed, when $a\rightarrow 0$, the shadow becomes a circle with
radius $3\sqrt{3}M/r_O$ as for the Schwarzschild black hole. When
we turn the spin on, the shadow shifts to the right (the spin
direction points to the top in Fig. \ref{Fig shadow}) and becomes
more compressed. With our coordinate system, photons detected on
the left side of the black hole have rotated in the same sense of
the black hole whereas photons detected on the right side have
counter-rotated. As mentioned before, prograde photons are allowed
to get closer to the black hole (up to $\bar r_+$) and therefore
reach the observer with a small angular separation from the black
hole. In the extremal limit $a\rightarrow M$, the shadow border on
the left (prograde) side is flattened while $\bar r_+$ equals the
horizon radius $r_H$. Beyond this limit, the horizon disappears
and the Kerr metric describes a naked singularity whose appearance
is no longer determined by the last circular orbits of the photons
but depends on the physical interpretation of the singularity
itself \cite{DeV}.

The idea of measuring the spin and other characteristics of
astrophysical black holes from the shape of the shadow border
dates back to Bardeen in 1972 \cite{Bar} and has been re-proposed
several times \cite{Cha,DeV,Shadow}. However, the real appearance
of an astrophysical black hole is largely dictated by the
properties of the accretion flow, as we shall discuss in the next
section.

\subsection{Caustics} \label{Sec Caustics}

As customary in gravitational lensing, the general functioning of
a lens model is determined by its caustic structure. In fact, by
studying caustics, we can find the points in which a source is
maximally amplified and the regions in which new pairs of images
are created. As showed in the previous section, spherically
symmetric black holes have degenerate caustics in the form of
zero-size tubes lying on the observer-lens axis behind and in
front of the black hole. It can be easily imagined that the loss
of spherical symmetry makes these tubes shift and acquire a finite
extension. The existence of finite-size caustics in the Kerr
metric was pointed in 1972 by Cunningham and Bardeen
\cite{CunBar}, who studied gravitational lensing of a star
orbiting a Kerr black hole. They found that additional pairs of
images were created when the source was in some particular
regions. The complete shape of the primary caustic was instead
illustrated for the first time by Rauch and Blandford
\cite{RauBla} who showed that it becomes a finite-size tube with
astroid cross-section, winding around the horizon. Lately, we
discussed the complete caustic structure of the Kerr metric,
illustrating higher order caustics and their dependence on spin
and the observer position \cite{CauKer}.

In order to calculate caustics, it is necessary to put Eqs.
(\ref{Geod1}) and (\ref{Geod2}) in the form of a lens mapping
relating the source coordinates\footnote{The source radial
coordinate $r_S$ is considered as a parameter. For each value of
$r_S$ we then obtain a section of the caustic tube at fixed
distance from the black hole.} $(\vartheta_S,\phi_S)$ to the image
position $(\theta_1,\theta_2)$. After that, we have to calculate
the Jacobian determinant and find the locus defined by its
vanishing (critical curve). Finally, the critical curve is mapped
to the source coordinates through the lens mapping to obtain the
caustic. Everything can be done numerically, but the details are
quite cumbersome and are summarized very shortly in this
treatment. On the other hand, some analytic formulae can be
obtained in the WDL and SDL analogously to what happens in
spherically symmetric black holes.

In brief, it is convenient to replace the coordinates in the
observer's sky $(\theta_1,\theta_2)$ by pseudo-polar coordinates
centered on the black hole position. Fixing the angular coordinate
$\eta$, the angular distance $\epsilon$ from the shadow, defined
in analogy to what is done in the SDL, determines how much the
photon is deflected from the black hole. In fact, a photon
detected far from the black hole ($\epsilon \gg 1$) is weakly
deflected, whereas the deflection diverges when the photon is
detected by the observer infinitely close to the shadow ($\epsilon
\rightarrow 0$). The number of oscillations in the polar angle
$\vartheta$ depends on the radial coordinate $\epsilon$. It turns
out that in the range $[\epsilon_m,\epsilon_{m+1}]$ in which the
photon performs $m$ inversions in the polar motion, there is only
one point in which the Jacobian determinant vanishes. Therefore,
we can identify the caustic order with the number of inversions in
the polar motion of photons that form the corresponding critical
images. The primary caustic is formed by photons experiencing only
one inversion in the polar motion. The secondary caustic is formed
by photons performing two inversions in the polar motion and so
on.

\begin{figure*}
\resizebox{\hsize}{!}{\includegraphics{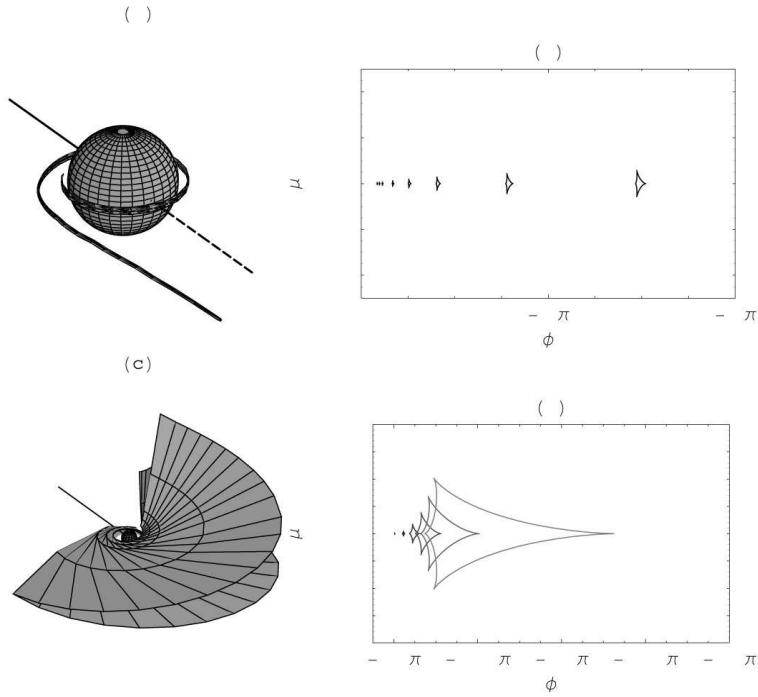}} \caption{(a) The
primary caustic tube for $a=0.9998M$ and $\mu_O=0$ in a
3-dimensional representation, with pseudo-euclidean coordinates
$x=r \sin \vartheta \cos \phi$, $y=r \sin \vartheta \sin \phi$,
$z=r \cos \vartheta$. The spin axis is directed toward the top.
The straight solid line indicates the direction towards the
observer, whereas the dashed line points in the opposite
direction. The primary caustic for a Schwarzschild black hole
($a=0$) coincides with this dashed line. (b) Cross section of the
same primary caustic at various distances. From left to right, the
radial coordinate is $r=5M, 4.5M, 4M, 3M, 2.3M, 1.82M, 1.42M,
1.22M$. (c) 3-dimensional picture of the second order caustic
surface of an extremal Kerr black hole for an observer on the
equatorial plane in the direction indicated by the solid line. The
surface has been plotted for radial distances in the range
$[2.2M,20M]$. (d) Second order asymptotic caustic for an
equatorial observer $\mu_o=0$ and different values of the spin.
From left to right, $a=0.02M,0.2M,0.4M,0.6M,0.8M,0.9998M$.}
 \label{Fig Cau}
\end{figure*}

Fig. \ref{Fig Cau}a is a 3-dimensional representation of the
primary caustic \cite{RauBla} for a quasi-extremal Kerr black hole
and Fig. \ref{Fig Cau}b shows its cross-section at various
distances. We can see that the zero-size tube of the Schwarzschild
black hole here becomes a tube with astroid cross-section winding
clockwise around the horizon an infinite number of times
\cite{RauBla,CauKer}. At large distances ($r_S \gg M$), the
caustic can be described analytically in the WDL, since the
photons of the critical images are weakly deflected. As shown by
Sereno and De Luca \cite{SerDeL}, the primary caustic can be
written in the coordinates $(\mu_S=\cos \vartheta_S, \phi_S)$ as
\begin{eqnarray}
\phi_S & = & -\pi  -4 a_m \varepsilon^2 -\frac{5}{4}\pi a_m \varepsilon^3  \nonumber \\
 & + &
\left[ \left( \frac{225}{128}\pi^2 -16 \right) a_m -\frac{15}{16}
\pi a_m^2 \sqrt{1-\mu_O^2}
\cos^3 \eta  \right] \varepsilon^4 \label{WFcaus1}  \\
\mu_S & = & -\mu_O  -\frac{15}{16} \pi a_m^2
(1-\mu_O^2)^{3/2}\varepsilon^4 \sin^3 \eta  \label{WFcaus2},
\end{eqnarray}
with $a_m=a/M$, $\eta$ ranging from $-\pi$ to $\pi$ and
$\varepsilon$ being the WDL expansion parameter already introduced
in spherically symmetric black holes in Eq. (\ref{varepsilon}). It
is interesting to note that the caustic remains infinitely thin at
second and third order in $\varepsilon$. We need to push the
expansion to the fourth order in order to find a finite extension
with the typical astroid cross-section. This fact signals that to
lowest order the Kerr black hole can be perfectly mimicked by a
Schwarzschild black hole with a finite shift, as previously noted
by Asada and Kasai \cite{AsaKas}. The angular size of the primary
caustic varies with distance as $\varepsilon^4\sim r_S^{-2}$.
Therefore its physical extension shrinks as $r_S^{-1}$ far from
the black hole, severely limiting chances of observations of spin
effects in the WDL. The effects of the spin becomes sizeable at
stronger deflections, which can be obtained either with sources at
distances from the black hole comparable with the Schwarzschild
radius or considering higher order images.

The secondary caustic involves photons with two inversions in the
polar motion. In the Schwarzschild limit it reduces to an
infinitely thin tube on the same side of the observer. Its
corresponding critical curve is the first retro-lensing Einstein
ring. In Fig. \ref{Fig Cau}d we can see what happens when we turn
the spin on \cite{CauKer}. The secondary caustic becomes very
large and for quasi-extremal spins has an angular extension larger
than $\pi$ in the azimuthal direction. Furthermore, the angular
size does not decrease with the distance $r_S$, and the whole
cross-section tends to an asymptotic astroid shape that is what we
show in \ref{Fig Cau}d. An impressive 3-dimensional view of the
secondary caustic in the quasi-extremal case is shown in Fig.
\ref{Fig Cau}c. The caustic surface includes most of the
equatorial plane assuming a discoidal shape. Creation and
destruction of images is thus much more common in the second
Einstein ring, than in the primary one. The spin effects, which
are crucial to determine the size and the position of the
secondary caustic, are then much more evident on higher order
images, but we must be aware that these are much fainter than the
primary images\footnote{Fig. \ref{Fig mag} in the Schwarzschild
case may be considered as a good reference even in the Kerr case
for a rough estimate of the luminosity of images of various
orders.}. Also for higher order caustics it is possible to derive
analytical approximations in a specific limit (SDL and small
values of the spin, with $r_S$ free to vary from $r_H$ to
infinity) \cite{SDLDLS,BozKer}. In this case for a caustic of
order $m\geq 2$ we have
\begin{eqnarray}
&& \mu_S= (-1)^m\left[\mu_O + R_m(1-{\mu_o}^2)^{3/2}\sin^3 \eta
\right],
\label{CauSDL1} \\
&& \phi_S=-m\pi-\Delta \phi_m+R_m\sqrt{1-{\mu_O}^2} \cos^3 \eta,
\label{CauSDL2}
\end{eqnarray}
where the azimuthal shift is
\begin{equation}
\Delta
\phi_m=-a\left\{\frac{2m\pi}{3\sqrt{3}}+2\log\left(2\sqrt{3}-3\right)
+  \log\left[
\frac{(2\sqrt{r_{S}}+\sqrt{6M+r_{S}})}{3\sqrt{(r_{S}-2M)}}\right]
\right\}, \label{Shift}
\end{equation}
and the semi-amplitude of the caustic is
\begin{equation}
R_m=a^2
 \left[\frac{1}{18} (5m\pi+8\sqrt{3}-36)+\frac{\left(
9M+2r_{S}-2\sqrt{r_{S}}\sqrt{6M+r_{S}}\right)}{3\sqrt{3}\sqrt{r_{S}}\sqrt{6M+r_{S}}}\right].
\label{Size}
\end{equation}

As for the primary caustic, the shift is linear and the size is
quadratic in the spin. It should be noted that for small values of
the spin there is a degeneracy\footnote{The degeneracy is evident
when switching to coordinates centered on the zero-order caustic
position $\mu=(-1)^m \mu_O$. Then all quantities depend on the
combination $a \sqrt{1-\mu_O^2}$.} between the spin $a$ and the
latitude of the observer on the equatorial plane. The caustics
become smaller and smaller as the observer is moved from the
equatorial plane to the polar axis, finally returning to zero
shift and size when the observer is perfectly aligned with the
rotation axis restoring the axial symmetry. The degeneracy between
spin and observer latitude is only broken at high values of the
spin \cite{CauKer}.

\subsection{Formation of the images}

The discussion of the images in Kerr black hole lensing may start
from the Schwarzschild limit. In this limit, we know that we have
an infinite sequence of direct images and an infinite sequence of
indirect images. We have noticed that the image of order $m$ has
an azimuthal shift $\Delta\phi$ in the range $[(m-1)\pi,m\pi]$. In
the Kerr case, it is possible to maintain this identification
using the number of oscillations in the polar angle, instead
\cite{CunBar}.

Solving Eq. (\ref{Geod1}) explicitly in $\mu_S$, we have
\begin{equation}
\mu_S=\pm\mu_+ \mathrm{sn}\left(2K(k)\psi,k\right), \label{mus
solved}
\end{equation}
where $\mu_+$, $k$ and $\psi$ are functions of $J$ and $Q$,
$K(k)=F(\pi/2,k)$ is the complete elliptic integral of the first
kind and sn is the Jacobi elliptic function. As sn has period
$4K(k)$, $\psi$ counts the number of half-oscillations on the
equatorial plane. It turns out that there is always at least one
image in each interval $m-1<\psi<m$. More specifically, when the
source is inside the caustic of order $m$, there are three images,
otherwise only one.

Comparing with the Schwarzschild black hole, in that case we had
an image in each interval $(m-1)\pi<\Delta\phi<m\pi$. Direct
images were those with odd $m$ and indirect images were those with
even $m$. In the Kerr black hole the situation is similar, but the
azimuthal shift is no longer suitable to label them, because it
evolves differently for photons with different inclinations on the
equatorial plane. The argument $\psi$ of the sn function in Eq.
(\ref{mus solved}) provides the correct variable. Its explicit
expression is reported in Ref. \cite{CauKer}.

\begin{figure*}
\resizebox{\hsize}{!}{\includegraphics{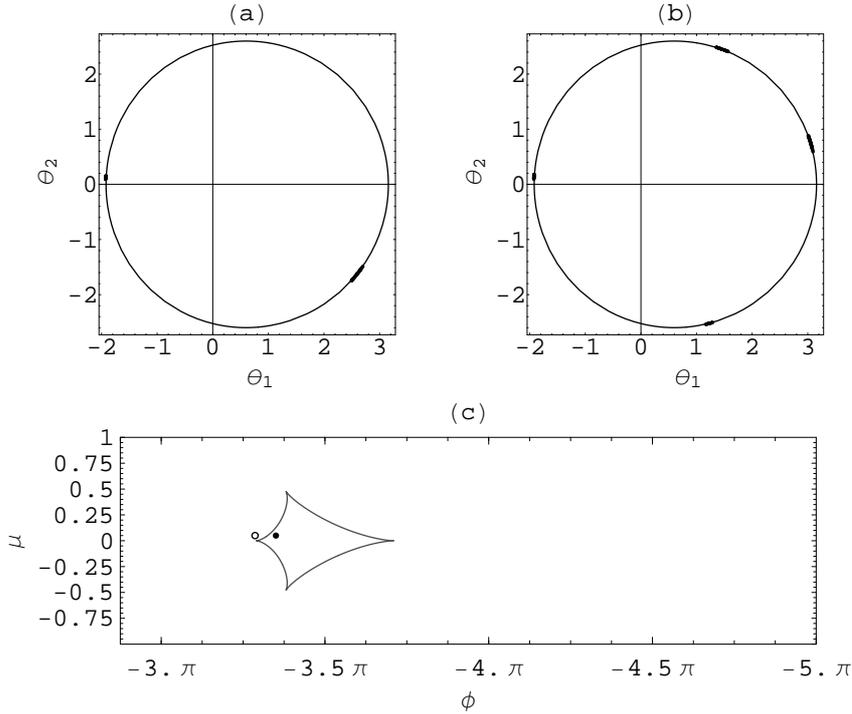}} \caption{Images
of order 3 and 4 of a source of radius $2M$ at $r_S=200M$ seen by
an observer on the equatorial plane of a black hole with spin
$a=0.6M$. In panel (c) we show two positions of the source with
respect to the third order caustic. In panel (a) we show the
images corresponding to the empty circle along with the shadow
border and in panel (b) we show the images corresponding to the
filled circle.}
 \label{Fig Ker Ima}
\end{figure*}

In Fig. \ref{Fig Ker Ima} we show the position of the images of
order 3 and 4 for a source very close to the caustic of order 3,
which involves photons performing a bit more than one loop and a
half  ($|\phi|\gtrsim 3\pi$). When the source is outside the
caustic (Fig \ref{Fig Ker Ima}a), there is one third order image
(on the left) and one fourth order image (on the bottom right).
The images appear as very stretched thin arcs. When the source is
inside the caustic, two more third order images appear on the top
right, while the fourth order image remains lonely.

The Kerr lens equation for sources close to caustics assumes a
very simple form for small values of the spin. In this limit, we
have \cite{BozRev,BozKer,SerDeL}
\begin{equation}
\delta \vartheta_s \cos \eta + (-1)^m (\delta \phi_s -R_m \cos
\eta)\sin \eta = 0, \label{Eqimages}
\end{equation}
where $(\delta \vartheta_s,\delta \phi_s)$ is the source position
relative to the center of the caustic, $R_m$ is the caustic
radius\footnote{It can be obtained from Eqs. (\ref{WFcaus1}) and
(\ref{WFcaus2}) for $m=1$ and from Eq. (\ref{Size}) for $m>1$.},
and $\eta$ is the angular coordinate around the shadow border in
the observer's sky as defined in Section \ref{Sec Caustics}. The
distance from the shadow $\epsilon$ remains the same as in
Schwarzschild to lowest order in $a$.

For near-extremal black holes ($a \rightarrow M$), the azimuthal
size of higher order caustics exceeds $2\pi$ \cite{CauKer}. As
$\phi$ is a periodic coordinate, this means that we should
consider the possibility that a source crosses the same caustic
surface at different number of loops. When this happens, new pairs
of additional images appear at the same order $m$ but with $\Delta
\phi$ differing by multiples of $2\pi$. From the physical point of
view, this means that these pairs of images are formed by photons
performing the same number of oscillations $m$ in the polar angle
but a different number of loops in azimuth $\phi$. These new pairs
of images appear on the flattened side on the left of the shadow
and are formed by prograde photons moving very close to the
equatorial plane. As the extension of higher order caustics
increases exponentially with the order $m$, the number of images
also grows exponentially with $m$ in extremal Kerr black holes.
However, the magnification drops even faster, so that it will be
extremely difficult to observe this effect.

\section{Observational perspectives}

Black holes offer a possibility (unique in the astrophysical
panorama) of studying light bending beyond the first order WDL
approximation, which governs all known cases of gravitational
lensing. However, unveiling this new phenomenology requires very
powerful technical capability. Let us start with a general
discussion.

The first corrections to the primary and secondary images, which
are well-known in all standard gravitational lensing situations,
scale with the combination $\theta_E\varepsilon$, which is
proportional to $M/r_O$. Interestingly, also the angular size of
the shadow of the black hole and all observables related to higher
order images are proportional to the ratio $M/r_O$. We shall
therefore take the shadow angular size $\bar
\theta=3\sqrt{3}M/r_O$ as the guiding quantity to evaluate
astrophysical black holes as candidates for gravitational lensing
beyond the WDL standard approximation.

Typical stellar black holes in the Galaxy have
\begin{equation}
\bar \theta_{Stellar}=5.1\times 10^{-11} \mathrm{arcsec}
\left(\frac{M}{M_\odot}\right) \left(\frac{r_O}{1\;
\mathrm{kpc}}\right)^{-1},
\end{equation}
which is definitely a too small angle to be observed. Much
promising candidates are massive black holes at the center of
galaxies. In particular, the black hole in the center of the Milky
Way, identified with the radio source Sgr A* \cite{Rich}, has mass
4.31 $\times 10^6 M_\odot$ \cite{Gillessen09} and lies at about 8
kpc from the sun. The angular size of its shadow is thus $\bar
\theta_{Sgr A*}=28 \mu$as. It is worth noting that some
supermassive black holes in other galaxies reach similar orders of
magnitude. For example, M87 has a central black hole of $3.2\times
10^9 M_\odot$ at a distance of 15 Mpc from us \cite{Macchetto},
resulting in a shadow radius of $11\mu$as. However, we must be
aware that the luminosity of the sources falls down as
$d_{OS}^{-2}$, limiting possible gravitational lensing
observations. The black hole in the center of our Galaxy was
proposed as the best candidate for black hole lensing by
Cunningham and Bardeen \cite{CunBar}. Most of the works in this
field have thereafter focussed on this nearby massive black hole.

Sgr A* has been intensively studied in the radio and X bands,
showing the existence of daily flares coming from the inner
accretion regions and probably generated by clumps of matter
falling down the horizon. As the Galactic center region is
optically thick, the near infrared band offers the best
opportunity to study stellar sources around Sgr A*. The latter has
only been detected in its flare state in the K-band at 15
magnitudes \cite{Genzel03}. We refer the interested reader to the
textbook by Melia \cite{Melia} for an extensive review summarizing
observations and models of Sgr A* and its environment.

Weak gravitational lensing by Sgr A* on background stars should be
quite common \cite{WarYus}, though it has never been identified
because of the crowded environment. Unfortunately, spin effects
are negligible on the primary and secondary images in the WDL
regime. This can be deduced by the size of the primary caustic,
which shrinks to zero at large distances. In the case of Sgr A*,
the size of the primary caustic is only 1170 km at
$r_S=100R_{Sch}$. It is therefore necessary to consider either
sources close to the black hole or higher order images, which are
however very faint and appear very close to the shadow of the
black hole within its accretion disk \cite{VirEll,Micro}. In the
following subsections, we will discuss several observational
proposals which aim at detecting effects beyond the basic WDL in
gravitational lensing by Sgr A* or other black holes.

\subsection{Sources orbiting the black hole}

Stars orbiting the black hole provide natural sources for
gravitational lensing. This possibility was first considered by
Cunningham and Bardeen \cite{CunBar}, who traced the apparent
trajectories in the sky of the primary and secondary images for a
star orbiting a black hole in a circular orbit. In Fig. \ref{Fig
CunBar} we see that the primary image is practically unaffected
when the source passes in front of the black hole. When the source
transits behind the black hole, the primary image moves around the
primary Einstein ring, designing a hat figure. Meanwhile, the
secondary image is opposite to the primary image slightly inside
the primary Einstein ring. However, when the source moves in front
of the black hole, the secondary image is closer to the black
hole, approaching the secondary critical curve. Very
interestingly, this picture shows the formation of two additional
second order images and the annihilation of one of these with the
original secondary image. These two events occur when the source
enters and exits the giant secondary caustic showed in Fig.
\ref{Fig Cau}c.

\begin{figure*}
\resizebox{\hsize}{!}{\includegraphics{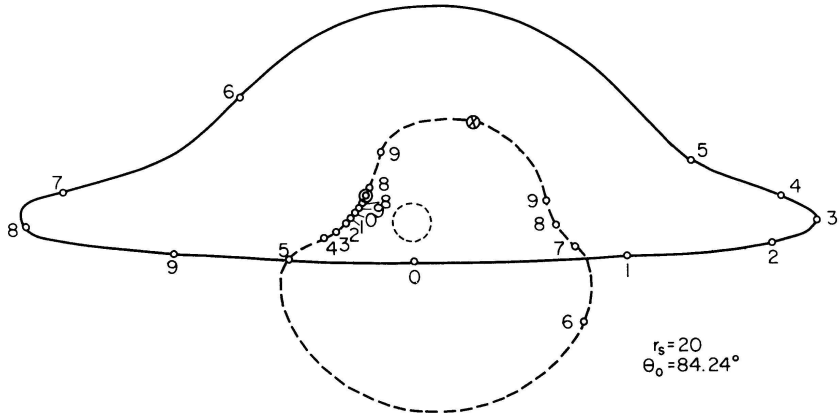}}
\caption{Apparent positions of the primary (solid line) and
secondary (dashed line) images of a source at $r_S=20M$ from an
extremal Kerr black hole. The observer is at a polar angle
$\vartheta_O=84.24^\circ$. Ticks mark the positions of the images
at 10 equally spaced times. As the source crosses the secondary
caustic, two new second order images are formed at the point
$\circledcirc$ and disappear at $\otimes$ when the source exits
from the caustic. The inner dashed circle has angular radius
$M/r_O$. Picture taken from Cunningham and Bardeen (1973)
\cite{CunBar}.}
 \label{Fig CunBar}
\end{figure*}

\begin{figure*}
\resizebox{8cm}{!}{\includegraphics{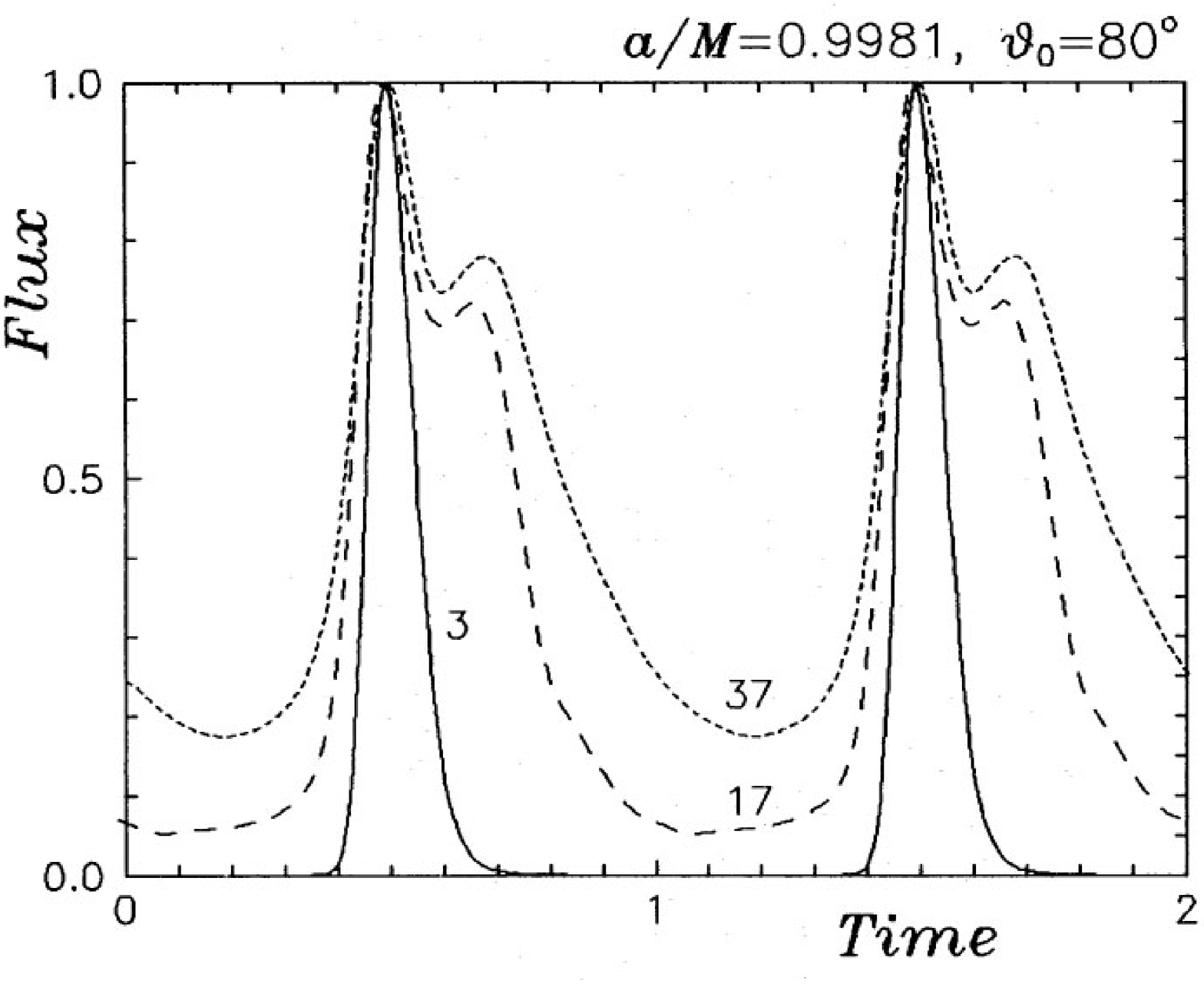}} \caption{Light
curve of a bright spot orbiting a quasi-extremal Kerr black hole
at different radii as indicated in the curve labels. Time is
normalized to the period. Picture taken from Karas, Vokrouhlicky
and Polnarev (1992) \cite{KVP}.}
 \label{Fig KVP}
\end{figure*}

We do not know of any stars on a circular equatorial orbit so
close to Sgr A*. Yet, the picture of Fig. \ref{Fig CunBar} may be
likely applied to a bright spot of hot material lying on the
accretion disk. The existence of such features has been proposed
by Abramovicz et al. in order to model the rapid X-ray variability
of Active Galactic Nuclei (AGN) \cite{Abramovicz}. Substantial
amount of work on this possibility has been developed thereafter
\cite{RauBla,KVP,Spot}. Assuming that we do not have enough
angular resolution to see the distinct images, the flux of the hot
spot would be periodically modulated through gravitational lensing
by the black hole. As the spot would rotate very quickly in a
strong gravitational field environment, Doppler and gravitational
red-shift cannot be neglected in a correct treatment. Note that if
the disk is opaque to the electromagnetic radiation considered,
only the primary image contributes to the observed flux, whereas
the second order and all higher order images are stopped by the
disk. In Fig. \ref{Fig KVP}, we see that the light curve is
dominated by a peak occurring when the spot is close to the
primary caustic behind the black hole. This peak may be followed
by a secondary peak, depending on the spot distance, caused by
Doppler enhancement.

Gravitational lensing of hot spots is very involved because of the
large number of parameters to be guessed, such as the size, the
shape, the position and the internal velocities of the spot,
besides the disk influence on the electromagnetic radiation
emitted by the spot. On the other hand, we know of many stars
orbiting Sgr A* at larger distances on eccentric orbits
\cite{Gillessen09,S-stars}. Spectra of these stars show that they
are early-type unexpectedly young stars; thus, their origin is
still under debate. If lensed, these stars would provide much
cleaner data \cite{S-stars lensing}. The best candidate is S6,
whose secondary image reaches $K=20.8$, with an angular separation
of 0.3 mas from the central black hole \cite{S6}. This is just
beyond the limits of the forthcoming interferometric instrument
GRAVITY at VLT \cite{GRAVITY}.

Another interesting possibility comes from Low Mass X-Ray Binaries
(LMXB) orbiting Sgr A* \cite{Muno}. In these objects, the X-ray
radiation is emitted from a very small region, providing good
candidates for probing the caustic structure of the black hole in
the case it has a non-negligible spin \cite{BozKer}.

\subsection{Broad Iron lines in the X-ray spectrum}

Among the known manifestations of gravitational lensing by black
holes, the broadening of the iron lines is the one that has
attracted more interest, as it is accessible with present
technology. This phenomenon was first discovered in the Seyfert 1
galaxy MCG-6-30-15 by Tanaka et al. \cite{Tanaka}, in which a
bright emission line is present over the X continuum at 6.4 keV.
This line has a characteristic doubly horned structure, with the
blue horn higher than the red horn.

This shape has been interpreted as follows: cold material in the
accretion disk is ionized by radiation from hot material
surrounding it. The following recombination is responsible for the
emission line, whose broad structure is due to the strong rotation
of the disc, which Doppler shifts the emission. The two horns come
from material whose velocity is directed along (or closely to) the
line of sight in both ways, with the blue horn appearing more
intense thanks to Doppler boosting. This picture also applies to
several lines at lower energies, which are generated far from the
black hole. The specifity of the iron lines comes from the fact
that they probe very high energies. In particular, the $K_\alpha$
line at 6.4 keV comes from electrons falling directly to the
K-shell. Such energies are only reached in regions very close to
the black hole (few Schwarzschild radii). At such distances,
general relativistic effects cannot be neglected in the
description of the line profiles and the whole spectrum in general
\cite{Cun1975}.

The iron $K_\alpha$ line has been observed in several AGNs and the
number of works attempting a numerical or analytic description of
relativistic iron line profiles has flourished. The first
treatments in the Schwarzschild metric have been done in Refs.
\cite{IronSch}. Lines in the Kerr metric have been studied in
Refs. \cite{IronKerr}. Usually, only the primary image of the disk
fluorescence is taken into account, because the accretion disk is
supposed to be thick. However, the second order image is typically
visible if the disc is seen nearly edge-on. The relevance of
higher order images has been pointed out in Ref. \cite{BHO} and
later in Ref. \cite{Fanton}. A non-exhaustive reference list for
later studies is in Ref. \cite{IronOthers}. Recently, also the
L-line at 0.9 keV, coming from electrons absorbed to the L-shell,
has been observed \cite{L-line}.

\begin{figure*}
\resizebox{12cm}{!}{\includegraphics{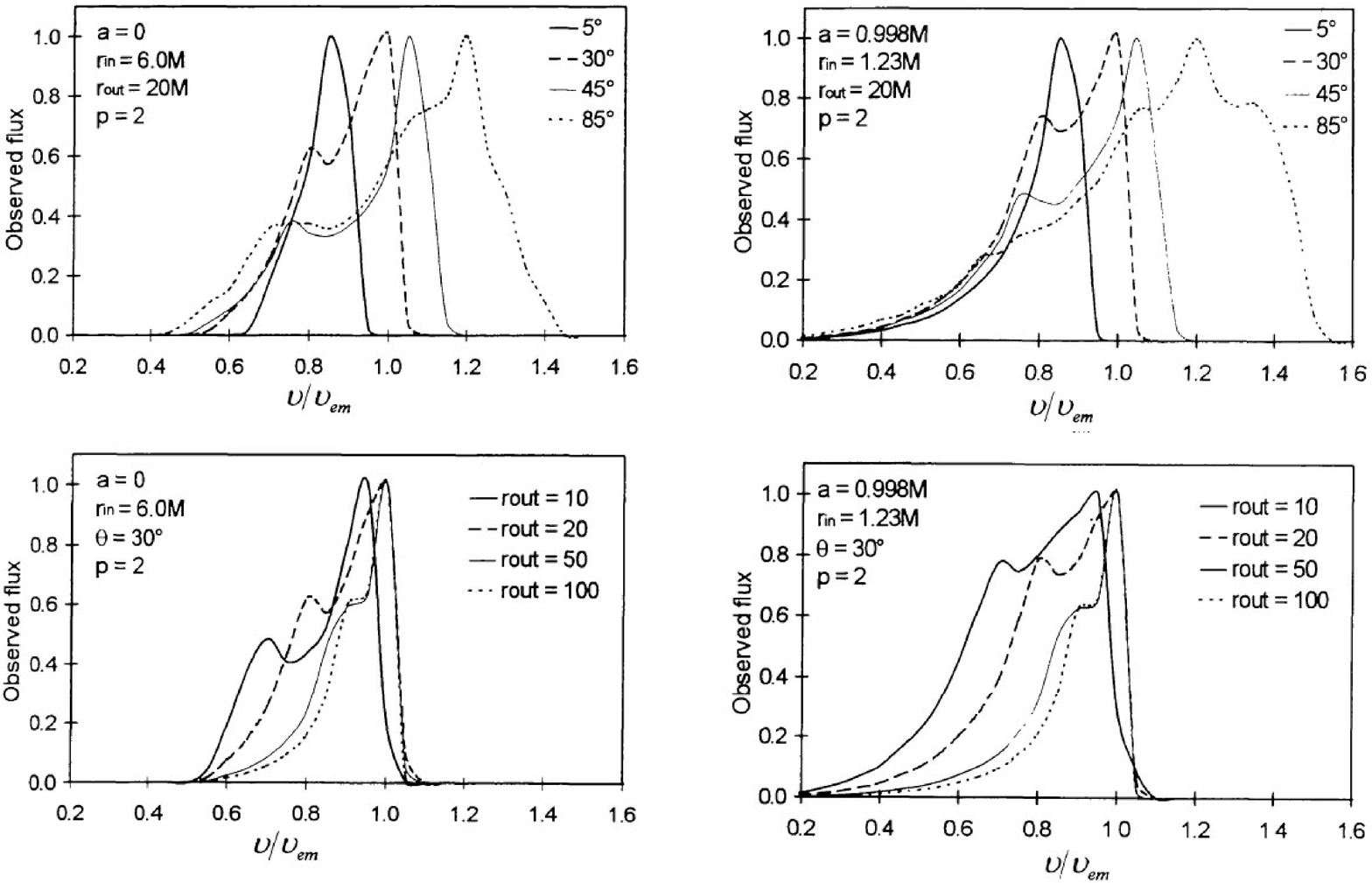}}
\caption{Relativistic line profiles for emitting disks around a
Schwarzschild (left panels) and a quasi-extremal Kerr (right
panels) black hole. In the top panels, line profiles for various
viewing angles $\vartheta_O$ are shown. In the bottom panels, the
outer radius of the disk $r_{out}$ is varied. Picture taken from
Fanton et al. (1997) \cite{Fanton}.}
 \label{Fig Fanton}
\end{figure*}

Fig. \ref{Fig Fanton} shows some relativistic line profiles
calculated by Fanton et al. \cite{Fanton}. The lines are broader
for a disc seen edge-on, showing up to four subpeaks. The line
returns single-peaked and is slightly redshifted for a face-on
disc. The lines are broader in a Kerr black hole, but there is
almost no difference if the inner radius of the emitting region is
larger than $6M$.

Models of relativistic line profiles depend on many parameters and
assumptions on the underlying model, such as opacity, thickness,
homogeneity, outer and inner radius of the emitting region,
emissivity profile and more and more. It is therefore difficult to
use them to extract the black hole parameters with confidence.
Nevertheless, they witness the existence of accretion disks at few
Schwarzschild radii of distance from the black hole and, at
present, they represent the cleaner information we can get from
these mysterious regions.

\subsection{Direct imaging}

Direct imaging of a black hole with a resolution of the order of
the angular size of the Schwarzschild radius is one of the ancient
dreams of astrophysicists. Up to now, we have always dealt with
indirect information on the regions surrounding the horizon. For
example, the measured spectra are the superposition of signals
coming from different regions experiencing different physics, from
the interstellar environment to the accretion flow up to possible
jets. Direct imaging would disentangle these signals and allow a
separate study of the different components, unveiling the secrets
of the black hole environment.

The first studies on the realistic appearance of a Schwarzschild
black hole surrounded by a Keplerian disk date back to Luminet
\cite{Lum}, who showed how the disk on the prograde side of the
black hole appears brighter thanks to the Doppler boost (which was
also responsible of the asymmetry in the iron line profiles, as we
have seen in the previous subsection). Fukue and Yokoyama drew
pictures at different wavelengths \cite{FukYok}. Viergutz
considered a vertically extended disk around a Kerr black hole
\cite{Vie}, whereas the first pictures of the realistic appearance
of a Kerr black hole were presented by Fanton et al.
\cite{Fanton}. Afterwards, Falcke, Melia and Agol presented
pictures of spherical inflow models taking into account the
electron scattering in the mm band \cite{FMA}. Polarimetric images
were presented in Ref. \cite{BML}. A non-exhaustive list of later
references is in Ref. \cite{Appearance}.

\begin{figure*}
\resizebox{12cm}{!}{\includegraphics{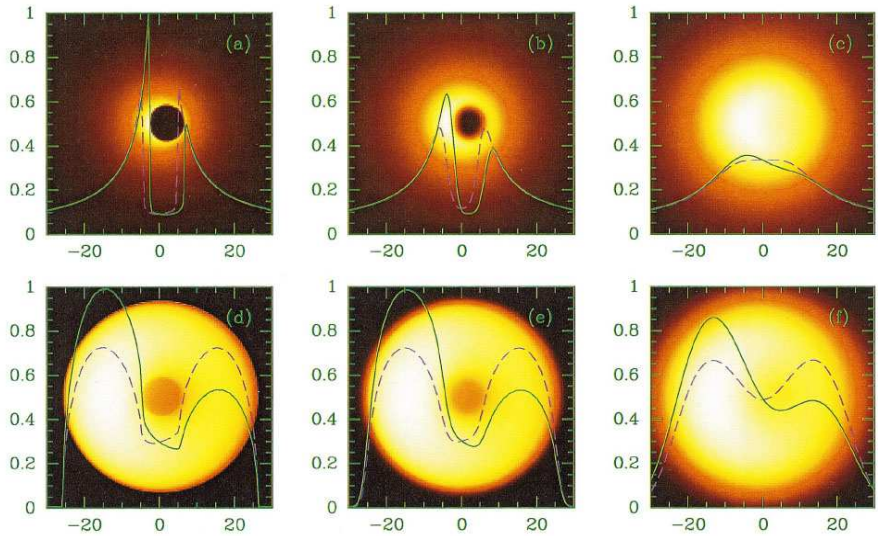}} \caption{Top
panels: extremal Kerr black hole with radial inflow. Bottom
panels: Schwarzschild black hole with flow on Keplerian shells.
Left panels: unscattered plots. Middle panels: the same plots
including scattering by interstellar medium at $\lambda=0.6$ mm.
Right panels: the same plots at $\lambda=1.3$ mm. Picture taken
from Falcke, Melia and Agol (1999) \cite{FMA}.}
 \label{Fig FMA}
\end{figure*}

In Fig. \ref{Fig FMA} we report the pictures by Falcke, Melia and
Agol \cite{FMA}. The difference between Kerr and Schwarzschild is
mostly in the shape of the shadow, which is flattened on the
prograde side for Kerr. The prograde side of the flow is
Doppler-boosted in both cases, giving poor indication on the spin
of the black hole itself. We can also appreciate the blurring
caused by scattering on electrons in the interstellar medium. Very
Long Baseline Interferometry (VLBI) observations in the cm band
show that the structure of Sgr A* is washed out by scattering,
with its apparent size scaling with the wavelength as $\lambda^2$
\cite{Lo}. Its intrinsic structure should be accessible at sub-mm
wavelengths, on which Ref. \cite{FMA} focussed.

The best angular resolution so-far achieved by human
instrumentations has been reached at $\lambda=3$ mm by VLBI on an
intercontinental scale, with 18 $\mu$as \cite{Kri02}. Even if this
is less than the apparent shadow of Sgr A*, interstellar
scattering is still too high at this wavelength. Observations of
M87 at this wavelength have also been carried out \cite{Kri06}.
This has inspired the idea that the black hole in M87 could be an
even more interesting target than Sgr A* \cite{AppearanceM87},
thanks to its jet structure. Meanwhile, projects for pushing the
global network of VLBI to shorter wavelengths are under discussion
\cite{Kri08} and represent the most promising frontier for direct
imaging of black holes. Indeed, present observations at
$\lambda=1.3$ mm seem to point that the intrinsic size of the
emitting region of Sgr A* is smaller than the shadow size
\cite{Sgr structure}. This is compatible with a black hole
surrounded by an accretion disk whose prograde side luminosity is
enhanced by Doppler boost (see Fig. \ref{Fig FMA}) \cite{DAF}.

Dynamical features such as hot spots rotating around the black
hole may be directly imaged in the future \cite{Imaging Hotspots}.
These could also be detected at worst resolutions by following the
motion of the centroid of Sgr A* \cite{Centroid}.

In the near infrared bands, the best resolution is now obtained by
the Keck interferometer and the VLTI. The latter will be upgraded
with the new instrument GRAVITY \cite{GRAVITY}, which will take
the resolution to 3 mas in the K-band. This is far from the shadow
scale of the black hole but is a great advance for the
determination of stellar orbits, which put the tighter constraint
to the mass of Sgr A*. It is also important for the study of
flares in the infrared.

Finally, the MAXIM project for developing X-ray interferometry on
spacecraft has been under study for several
years\footnote{http://maxim.gsfc.nasa.gov.}. The incredible
difficulties posed by this challenge will be rewarded by a black
hole imager with resolution even better than 1 $\mu$as. The new
frontiers opened by such a futuristic instrument can hardly be
imagined.

\begin{acknowledgements}
I wish to thank Mauro Sereno, Luigi Mancini, Arlie Petters and
Eric Agol for useful suggestions and comments on the manuscript. I
acknowledge support by Agenzia Spaziale Italiana, by PRIN (Prot.
2008NR3EBK\_002), and by research funds of Salerno University.

\end{acknowledgements}


\begin{thebibliography}{}
%
%\bibitem{KliFri} T. Kling and S. Frittelli, ApJ, 675, 115
%(2008).

\bibitem{Dar}  C. Darwin, Proc. of the Royal Soc. of London A, 249, 180
(1959).

\bibitem{Wei} S. Weinberg, "Gravitation and Cosmology", John Wiley \&
Sons, New York (1972).

\bibitem{CVE} C.-M. Claudel, K.S. Virbhadra and G.F.R. Ellis, Jour. of Math. Phys., 42, 818 (2001).

\bibitem{Atk} R. Atkinson, Astron. J, 70, 517 (1965).

\bibitem{Oha} H.C. Ohanian, Am. J. Phys., 55, 428 (1987).

\bibitem{VirEll} K.S. Virbhadra and G.F.R. Ellis, Phys. Rev. D, 62, 084003
(2000).

\bibitem{BCIS} V. Bozza, S. Capozziello, G. Iovane, and G. Scarpetta, Gen.
Rel. and Grav., 33, 1535 (2001).

\bibitem{FriNew} S. Frittelli and E.T. Newman, Phys. Rev. D, 59,
124001 (1999).

\bibitem{FKN} S. Frittelli, T.P. Kling, and E.T. Newman, Phys. Rev. D, 61,
064021 (2000).

\bibitem{Perlick} V. Perlick, Phys. Rev. D, 69, 064017 (2004).

\bibitem{LivRev} V. Perlick, Living Rev. Relativity, 7, 9
(2004).

\bibitem{Nem} R. Nemiroff, Am. Jour. Phys., 61, 619 (1993).

\bibitem{ComLEq} V. Bozza, Phys. Rev. D, 78, 103005 (2008).

\bibitem{Lum} J.P. Luminet, A\&A, 75, 228 (1979).

\bibitem{HolWhe} D.E. Holz and J.A. Wheeler, ApJ, 578, 330 (2002).

\bibitem{DecFol} Y. D\'ecanini and A. Folacci, Phys. Rev. D, 81, 024031
(2010).

\bibitem{EpsSha} R. Epstein and I. Shapiro, Phys. Rev. D, 22, 2947 (1980);

\bibitem{WDLSph} E. Fischbach and B.S. Freeman, Phys. Rev. D, 22, 2950 (1980); G.
Richter and R. Matzner, Phys. Rev. D, 26, 1219 (1982); 26, 2549
(1982); A.F. Sarmiento, Gen. Rel. and Grav., 14, 793 (1982); M.
Sereno, Phys. Rev. D, 67, 064007 (2003).

\bibitem{WDL} C. R. Keeton and A. O. Petters, Phys. Rev. D 72, 104006
(2005); 73, 044024 (2006); Phys. Rev. D 73, 104032 (2006).

\bibitem{Ebina}
 J. Ebina, T. Osuga, H. Asada, and M. Kasai, Prog. of Th. Phys., 104,
 1317 (2000); G.F. Lewis, X.R. Wang, Prog. of Th. Phys., 105, 893
 (2001).

\bibitem{SerRei} M. Sereno, Phys. Rev. D,
69, 023002 (2004).

\bibitem{DamEsp} Th. Damour and G. Esposito-Far\`ese, Phys. Rev. D,
53, 5541 (1996).

\bibitem{SerCos} M. Sereno, Phys. Rev. D,
77, 043004 (2008).

\bibitem{Cha} S. Chandrasekhar,``Mathematical Theory of Black Holes'',
Clarendon Press, Oxford (1983).

\bibitem{ReiNor} E.F. Eiroa, G.E. Romero and D.F. Torres, Phys. Rev.
D, 66, 024010 (2002).

\bibitem{Boz1} V. Bozza, Phys. Rev. D, 66, 103001 (2002).

\bibitem{EirTor} E.F. Eiroa and D.F. Torres, Phys. Rev.
D, 69, 063004 (2004).

\bibitem{SDLDLS} V. Bozza and G. Scarpetta, Phys. Rev. D, 76, 083008 (2007).

\bibitem{Spheric} A. Bhadra, Phys. Rev. D 67, 103009 (2003); E. F.
Eiroa, Phys. Rev. D 71, 083010 (2005); R. Whisker, Phys. Rev. D
71, 064004 (2005); A. S. Majumdar and N. Mukherjee, Int. J. Mod.
Phys. D 14, 1095 (2005); Eiroa, Phys. Rev. D, 73, 043002 (2006);
K. K. Nandi, Y.-Z. Zhang, and A. V. Zakharov, Phys. Rev. D, 74,
024020 (2006);  K. Sarkar and A. Bhadra, Class. Quant. Grav. 23,
6101 (2006); N. Mukherjee, A.S. Majumdar, Gen. Rel. and Grav., 39,
583 (2007); F. Rahaman, M. Kalam and S. Chakraborty, Chin. J.
Phys., 45, 518 (2007); T. K. Dey, S. Sen, Mod. Phys. Lett. A, 23,
953 (2008); S. Chen, J. Jing, Phys. Rev. D, 80, 024036 (2009); T.
Ghosh and S. Sengupta, Phys. Rev. D, 81, 044013 (2010).

\bibitem{BozRev} V. Bozza, Nuovo Cimento B, 122, 547 (2007).

\bibitem{IyePet} S.V. Iyer and A.O. Petters, Gen. Rel. and Grav., 39, 1563 (2007).

\bibitem{TimDel} V. Bozza and L. Mancini, Gen. Rel. and Grav., 36, 435 (2004).

\bibitem{ChaRef} K. Chang and S. Refsdal, Nat., 282, 561 (1979);
A\& A, 132, 168 (1984).

\bibitem{Poi} E. Poisson, Phys. Rev. Lett., 94, 161103 (2005).

\bibitem{BozSer} V. Bozza and M. Sereno, Phys. Rev. D, 73, 103004 (2006).

\bibitem{Amore} P. Amore and S. Arceo, Phys. Rev. D, 73, 083004 (2006);
P. Amore, S. Arceo, and F.M. Fernandez, Phys. Rev. D, 74, 083004
(2006); P. Amore, M. Cervantes, A. De Pace, and F.M. Fernandez,
Phys. Rev. D, 75,083005 (2007).

\bibitem{Car} B. Carter, Phys. Rev., 174, 1559 (1968).

\bibitem{RauBla} K.P. Rauch and R.D. Blandford, ApJ, 421,
46 (1994).
%
\bibitem{KVP} V. Karas, D. Vokrouhlicky and A.G. Polnarev, MNRAS,
259, 569 (1992).

\bibitem{Vie} S.U. Viergutz, A\&A, 272,
355 (1993).

\bibitem{Fanton} C. Fanton, M. Calvani, F. de Felice, and A. Cadez, PASJ, 49,
159 (1997).

\bibitem{DexAgo} J. Dexter and E. Agol, ApJ, 696, 1616 (2009).

\bibitem{IyeHan1} S.V. Iyer and E.C. Hansen, Phys. Rev. D, 80, 124023 (2009).

\bibitem{PinRoe} S. Pineault and R.C. Roeder, ApJ, 212, 541
(1977); 213, 548 (1977).

\bibitem{WDLKer} I. Bray, Phys. Rev. D, 34, 367 (1986); A. Edery and J. Godin, Gen. Rel. and Grav., 38, 1715
(2006); M.C. Werner and A. O. Petters, Phys. Rev. D, 76, 064024
(2007)

\bibitem{AsaKas} H. Asada and M. Kasai, Prog. Theor. Phys., 104, 95
(2000); H. Asada, M. Kasai, and T. Yamamoto, Phys. Rev. D, 67,
043006 (2003).

\bibitem{SerDeL} M. Sereno and F. De Luca, Phys. Rev. D, 74, 123009
(2006); 78, 023008 (2008).

\bibitem{BozEq} S.E. Vazquez and E.P. Esteban, Nuovo Cim.
B, 119, 489 (2004); V. Bozza, Phys. Rev. D, 67, 103006 (2003).

\bibitem{BozKer} V. Bozza, F. De Luca, G. Scarpetta, and M.Sereno, Phys. Rev. D 72, 083003 (2005);
V. Bozza, F. De Luca, and G. Scarpetta, Phys. Rev. D 74, 063001
(2006).

\bibitem{DeV} A. de Vries, Class. Quant. Grav., 17, 123 (2000); C.
Bambi and K. Freese, Phys. Rev. D 79, 043002 (2009).

\bibitem{Bar} J. Bardeen, in C. De Witt and B.S. De Witt, "Black Holes. \'Ecole d'\'et\'e de Physique Th\'eorique, Les Houches 1972". Gordon \& Breach Science Publishers, New York
(1973);

\bibitem{Shadow} R. Takahashi, ApJ, 611, 996 (2004); Publ. Astron. Soc. Japan, 57, 273, (2005); A.F. Zakharov et al., A\& A, 442, 795 (2005);
K. Hioki and K.-I. Maeda, Phys. Rev. D, 80, 024042 (2009).

\bibitem{CunBar}  C.T. Cunningham and J.M. Bardeen, ApJ 173, L137 (1972); ApJ {\bf 183},
237 (1973).

\bibitem{CauKer} V. Bozza, Phys. Rev. D, 78, 063014 (2008).

\bibitem{Rich} D. Richstone et al., Nature, 395, A14 (1998).

\bibitem{Gillessen09} S. Gillessen et al., ApJ, 692, 1075 (2009).

\bibitem{Macchetto} F. Macchetto et al., ApJ, 489, 579 (1997).

\bibitem{Genzel03} R. Genzel al., Nature, 425, 934 (2003).

\bibitem{Melia} F. Melia, ``The Black Hole at the center of our Galaxy'', Princeton University Press,
(2003).

\bibitem{WarYus} M. Wardle and F. Yusuf-Zadeh, ApJ, 387,
L65 (1992); M. Jaroszy\'nski, Acta Astron., 48, 413, (1998); T.
Alexander and A. Sternberg, ApJ, 520, 137 (1999); T. Alexander,
ApJ, 553, L149 (2001); J. Chanam\'e, A. Gould, and J.
Miralda-Escud\'e, ApJ, 563, 793 (2001); T. Alexander and A. Loeb,
ApJ, 551, 223 (2001); A. Nusser and T. Broadhurst, MNRAS, 355, L6,
(2004).

\bibitem{Micro} A.O. Petters, MNRAS, 338, 457 (2003).

\bibitem{Abramovicz} M.A. Abramovicz, A. Lanza, E.A. Spiegel, and E. Szuszkiewicz, Nature, 356,
41 (1992).

\bibitem{Spot} V. Karas and G. Bao, A\&A, 257, 531 (1992); G. Bao, A\&A, 257, 594
(1992); A.F. Zakharov, MNRAS, 269, 283 (1994); G. Bao, P. Hadrava,
and E. \O stgaard, ApJ, 425, 63 (1994); G. Bao and E. \O stgaard,
ApJ, 443, 54 (1995); Hollywood et al., ApJ, 448, L21 (1995); V.
Karas, ApJ, 470, 743 (1996); MNRAS, 288, 12 (1997); PASJ, 51, 317
(1999); J. Fukue, PASJ, 55, 1121 (2003).

\bibitem{S-stars} R. Sch\"odel et al., ApJ, 596, 1015 (2003); A.M. Ghez et al., ApJ, 620,
744 (2005).

\bibitem{S-stars lensing} F. De Paolis, A. Geralico, G. Ingrosso, and A.A. Nucita,
A\&A, 409, 809 (2003); V. Bozza and L. Mancini, ApJ, 611, 1045
(2004); 627, 790 (2005).

\bibitem{S6} V. Bozza and L. Mancini, ApJ, 696, 701 (2009).

\bibitem{GRAVITY} Eisenhauer et al., Proceedings of the SPIE, 7013,
70132 (2008).

\bibitem{Muno}  M.P. Muno et al., ApJ 622, L113 (2005).

\bibitem{Tanaka} Y. Tanaka et al., Nature, 375, 659 (1995).

\bibitem{Cun1975} C.T. Cunningham, ApJ, 202, 788 (1975); I. Asaoka, PASJ, 41, 763 (1989); M. Jaroszy\'{n}ski and A.
Kurpiewski, A\&A, 326, 419 (1997).

\bibitem{IronSch} A.C. Fabian, M.J. Rees, L. Stella, and N.E.
White, MNRAS, 238, 729 (1989); L. Stella, Nature, 344, 747 (1990);
A. Laor, ApJ, 376, 90 (1991).

\bibitem{IronKerr} Y. Kojima, MNRAS, 250, 629 (1991);
K. Chen and D. Eardley, ApJ, 382, 125 (1991); G. Bao, ApJ, 409,
L41 (1993), G. Matt, G.C. Perola, and L. Stella, A\&A, 267, 643
(1993); J.-M. Hameury, J.-A. Marck, and D. Pelat, A\&A, 287, 795
(1994).

\bibitem{BHO} G. Bao, P. Hadrava,
and E. \O stgaard, ApJ, 435, 55 (1994).

\bibitem{IronOthers} G. Bao, P. Hadrava,
and E. \O stgaard, ApJ, 464, 684; B.C. Bromley; K. Chen, and W.A.
Miller, ApJ; 475, 57 (1997); Ch.S. Reynolds, A.J. Young, M.C.
Begelman, and A.C. Fabian, ApJ, 514, 164 (1999); A. Martocchia, V.
Karas, and G. Matt, MNRAS, 312, 817 (2000); Q. Yu, Y. Lu, MNRAS,
311, 161 (2000); M. Dovciak, V. Karas, and T. Yaqoob, ApJS, 153,
205 (2004); K. Beckwith and Ch. Done, MNRAS, 352, 353 (2004); K.
Fukumura and S. Tsuruta, ApJ, 613, 700 (2004); B. Czerny et al.,
A\&A, 420, 1 (2004); A. Cade\v{z} and M. Calvani, MNRAS, 363, 177
(2005); R.W. Goosmann, A\&A, 454, 741 (2006); A.F. Zakharov and
S.V. Repin, New Astr., 11, 405 (2006); S.V. Fuerst and K. Wu,
A\&A, 474, 55 (2007); S.V. Repin, V.N. Lukash, and V.N. Strokov,
Astr. Reps., 52, 1 (2008).

\bibitem{L-line} A.C. Fabian et al., Nature, 459, 540 (2009).

\bibitem{FukYok} J. Fukue and T. Yokoyama, PASJ, 40, 15 (1988).

\bibitem{FMA} H. Falcke, F. Melia, and E. Agol, ApJ, 528, L13 (1999).

\bibitem{BML} B.C. Bromley, F. Melia, and S. Liu, ApJ, 555, L83 (2001);

\bibitem{Appearance} J. Fukue, PASJ, 55, 155 (2003); Takahashi, ApJ, 611, 996 (2004);
K. Beckwith and C. Done, MNRAS, 359, 1217 (2005); A.E. Broderick
and R. Narayan, ApJ, 638, L21 (2006); F. Yuan, Zh.-Q. Shen, and L.
Huang, ApJ, 642, L45 (2006); L. Huang, M. Cai, Zh.-Q. Shen, and F.
Yuan, MNRAS, 379, 833 (2007); Y.-F. Yuan, Z. Cao, L. Huang, and
Zh.-Q. Shen, ApJ, 699, 722 (2009).

\bibitem{Lo} K.Y. Lo et al., ApJ, 508, L61 (1998).

\bibitem{Kri02}  T.P. Krichbaum et al., Proc. 6th European VLBI Network Symposium, Bonn,
Germany (2002).

\bibitem{Kri06} T.P. Krichbaum et al., Jour. of Phys. Conf. Ser., 54, 328 (2006).

\bibitem{AppearanceM87} A.E. Broderick and A. Loeb, ApJ, 697, 1164
(2009).

\bibitem{Kri08} T.P. Krichbaum et al., arXiv:0812.4211

\bibitem{Sgr structure} Sh.S. Doeleman et al., Nature, 455, 78
(2008).

\bibitem{DAF} J. Dexter, E. Agol, and P. Ch. Fragile, APJ, 703,
L142 (2009).

\bibitem{Imaging Hotspots} A.E. Broderick and A. Loeb, MNRAS, 363, 353
(2005); 367, 905 (2006).

\bibitem{Centroid} A.E. Broderick and A. Loeb, ApJ, 636, L109
(2006).



\end{thebibliography}
\end{document}